\documentclass[twocolumn]{aastex63}
\usepackage{bm}
\usepackage{multirow}
\usepackage{amsmath}
\usepackage{color}
\usepackage[T1]{fontenc}

\newcommand{\hi}{{\textsc{Hi}}}

\newcommand{\mhi}{\ensuremath{\textrm{M}_{\textsc{Hi}}}}
\newcommand{\msun}{\ensuremath{\textrm{M}_{\odot}}}
\newcommand{\lgmhi}{\ensuremath{\log_{10}\mhi}}
\newcommand{\mstar}{\ensuremath{\textrm{M}_{\ast}}}
\newcommand{\lgmstar}{\ensuremath{\log_{10}\mstar}}
\newcommand{\nuvr}{\ensuremath{\textrm{NUV}-r}}
\newcommand{\qcr}{\fontfamily{qcr}\selectfont}

\shortauthors{Li et al.}

\begin{document}

\title{On the existence, rareness and uniqueness of quenched 
\hi-rich galaxies in the local Universe}

\correspondingauthor{Xiao Li \& Cheng Li}

\author[0000-0002-2884-9781]{Xiao Li}
\affiliation{Department of Astronomy, Tsinghua University, Beijing 100084, China}
\email{xli27938@gmail.com}

\author[0000-0002-8711-8970]{Cheng Li}
\affiliation{Department of Astronomy, Tsinghua University, Beijing 100084, China}
\email{cli2015@tsinghua.edu.cn}

\author[0000-0001-5356-2419]{H. J. Mo}
\affiliation{Department of Astronomy, University of Massachusetts Amherst, MA 01003, USA}

\author{Jianhong Hu}
\affiliation{Department of Astronomy, Tsinghua University, Beijing 100084, China}

\author[0000-0002-6593-8820]{Jing Wang}
\affiliation{Kavli Institute for Astronomy and Astrophysics, Peking University, Beijing 100871,  China}

\author[0000-0003-1938-8669]{Ting Xiao}
\affiliation{School of Physics, Zhejiang University, Hangzhou, Zhejiang 310058, China}

\begin{abstract}
Using data from ALFALFA, xGASS, HI-MaNGA and the Sloan Digital Sky Survey (SDSS),
we identify a sample of 47 ``red but \hi-rich''(RR) galaxies 
with \nuvr$~>5$ and unusually high \hi-to-stellar mass ratios. 
We compare the optical properties and local environments between
the RR galaxies and a control sample of ``red and \hi-normal''(RN) 
galaxies that are matched in stellar mass and color. 
The two samples are similar in the optical properties typical of massive 
red (quenched) galaxies in the local Universe. The RR sample tends to be 
associated with slightly lower-density environments and 
has lower clustering amplitudes and smaller neighbor counts at scales 
from several hundred kiloparsecs to a few megaparsecs. The results are consistent 
with the RR galaxies being preferentially located at the center of 
low-mass halos, with a median halo mass $\sim 10^{12}h^{-1}\msun$
compared to $\sim 10^{12.5}h^{-1}\msun$ for the RN sample.
This result is confirmed by the SDSS group catalog 
which reveals a central fraction of 89\% for the RR sample,
compared to $\sim$60\% for the RN sample. If assumed to follow
the \hi\ size-mass relation of normal galaxies, the RR 
galaxies have an average \hi-to-optical radius 
ratio of $R_{\rm HI}/R_{90}\sim4$, four times the average 
ratio for the RN sample. We compare our RR sample 
with similar samples in previous studies, and quantify 
the population of RR galaxies using the SDSS complete 
sample. We conclude that the RR galaxies form a unique but rare 
population, accounting for only a small fraction of the 
massive quiescent galaxy population. We discuss the formation 
scenarios of the RR galaxies.
\end{abstract}

\keywords{Neutral hydrogen clouds (690), Galaxy dark matter halos (1880)}

\section{Introduction} \label{sec:intro}

In current theory of structure formation, galaxies form at the center of
dark matter halos through gas cooling and condensation
\citep{1978MNRAS.183..341W,MoBoschWhite2010}.
As the dominant component of cold gas, the atomic 
hydrogen (\hi) reservoir is expected to play a vital role 
in regulating the rise and fall of star formation in 
galaxies. In fact, the \hi\ surface mass density 
in low-redshift galaxies has been found to tightly 
correlate with the surface density of their star formation 
rate (SFR), known as the Kennicutt-Schmidt law \citep{KSlaw}. 
In addition, large surveys of multiband imaging and 
spectroscopy as well as \hi\ emission at 21cm have
revealed strong correlations between the \hi-to-stellar 
mass ratio (\mhi/\mstar) and the optical and UV properties 
of galaxies,
with lower \hi\ mass fractions in galaxies with redder 
colors, lower SFRs and higher stellar mass densities
\citep[e.g.][]{Haynes1984,2004ApJ...611L..89K,2009MNRAS.397.1243Z,
GASS,HuangS2012,Li2012,Brown2015,xGASS,XiaoLi,Liu2023}. 
In particular, it has been widely accepted that 
the cessation of star formation in a galaxy must be 
associated with the reduction of its \hi\ reservoir. 
In a recent review, \citet{Saintoge2022} concluded that 
there is no evidence for a significant population of 
passive galaxies with \hi\ reservoirs comparable to those 
of star-forming galaxies, based on extensive analyses 
of the extended GALEX Arecibo SDSS survey 
\citep[xGASS;][]{GASS,xGASS}. 

On the other hand, given the large scatter
of the \hi\ scaling relations found in previous studies,
typically $0.2-0.3$ dex in $\log_{10}($\mhi/\mstar$)$,
one can expect that some galaxies will deviate significantly 
from the average relations. Studies of galaxies with 
\hi\ anomalies should in principle provide 
insights into the physical processes that drive gas-related 
star formation and quenching, and probably 
also cosmological models \citep[e.g.][]{Peebles2022}. 
In this regard, there has been a rich history of studies 
on the \hi\ content of early-type galaxies 
\citep[ETGs; e.g.][]{Knapp1985-ellipticals,Wardle1986-S0,
Bregman1992,Morganti2006,diSeregoAlighieri2007,Grossi2009,
Serra2012-ATLAS3D,Young2014}. The ATLAS$^{\rm3D}$ \hi\ survey 
\citep{Serra2012-ATLAS3D} obtained resolved \hi\ 
observations for a volume-limited sample of 166 nearby ETGs 
selected from ATLAS$^{\rm 3D}$ \citep{Cappellari2011-ATLAS3D}, 
detecting significant \hi\ in $\sim10\%$ of all the 
ETGs inside the Virgo cluster and $\sim40\%$ of those outside. 
In addition, the \hi\ discs of these galaxies are found 
to have much larger sizes than their optical parts, as well as 
much lower \hi\ column densities than what was found typically 
in spiral galaxies \citep{Serra2012-ATLAS3D,Serra2014}.
Although with large \hi\ detection rates, ETGs usually 
have low \hi\ mass fractions as expected from their relatively 
low SFRs. 


Some other attempts have been made to search for 
\hi\ content in E+A, or post-starburst (PSB) galaxies 
\citep[e.g.][]{Chang2001,Buyle2006-PSB,Buyle2008-PSB-merger,
Zwaan2013-PSB,Pracy2014-PSB},
a rare population of galaxies that are believed to have 
their star formation recently shut down, with no/little 
H$\alpha$ emission but strong high-order Balmer absorption
indicative of young stars. \citet{Zwaan2013-PSB} detected 
\hi\ emission in a high fraction (6/11) of E+A galaxies 
selected from the 2dF Galaxy Redshift 
Survey \citep[2dFGRS;][]{2dFGRS} and the Sloan Digital Sky Survey
\citep[SDSS;][]{2000AJ....120.1579Y}. A followup by 
\citet{Pracy2014-PSB} 
obtained integral field spectroscopy (IFS) for two E+A galaxies
with \hi\ emission, finding them to be consistent 
with a post-starburst population only in the central region,
with the outer regions of both galaxies presenting 
strong optical emission lines. It is thus not surprising
for a large fraction of the E+A galaxies in \citet{Zwaan2013-PSB}
to have \hi\ gas extending to large radii.

There have also been many observational studies to search 
``\hi-excess'' galaxies \citep[e.g.][]{2013MNRAS.433..270W,Huang2014,Lee2014,
Lemonias2014,Wang2015-Bluedisk,Gereb2016,Lutz2017,Gereb2018,
Lutz2018,Lutz2020,Randriamampandry2021,2022ARep...66..755Z}, 
which are usually strongly star-forming galaxies with late-type 
morphologies, identified as extreme outliers to the scaling relation
between \hi\ and stellar content of galaxies. Compared to galaxies with 
normal \hi\ contents,  these galaxies are found to have much larger \hi\ disks and lower \hi\ surface 
densities, but similar optical sizes and radial distributions 
of the gas-phase metallicity. \cite{Gereb2018} obtained \hi\ resolved 
maps and optical spectra for four \hi-excess galaxies, and 
found that the \hi\ disk and stellar bulge are 
counter-rotating in one of the galaxies, 
suggesting external origin of the excessive \hi. 
However, \citet{Lutz2020} found no indication for counter-rotation 
in any of their \hi-excess galaxies. By examining the star 
formation activity and \hi\ kinematics, \cite{Lutz2018} 
suggested that these galaxies stabilize their large \hi\ discs 
by their higher-than-average baryonic specific angular momentum,
likely inherited from the high spin of their host dark halos.

Several recent studies have focused on \hi-rich galaxies 
with suppressed SFRs 
\citep{Lemonias2014,Parkash2019,Guo2020,Wang2022-FAST,Sharma2023}. 
\citet{Lemonias2014} selected a parent sample of 258 
\hi-rich galaxies from 
the Arecibo Legacy Fast Arecibo L-band Feed Array Survey 
\citep[ALFALFA;][]{Giovanelli_2005,2018ApJ...861...49H},
defined as the top 5\% fraction in the distribution of the 
\hi\ mass-to-stellar mass ratio. From this parent sample,  
20 massive and low-sSFR galaxies with $\log_{10}$(\mstar/\msun)$~>10.6$, 
$\log_{10}({\rm sSFR})<-10.75$ and $b/a>0.5$ were selected for followup 
observations with the Very Large Array (VLA). 
\citet{Guo2020} analyzed a sample of 279 massive red 
spirals with \mstar$>10^{10.5}~$\msun\ and red colors 
defined by $u-r$, and found 74/166 in the ALFALFA
catalog to have \hi\ detections. \citet{Wang2022-FAST} 
obtained followup \hi\ observations with 
the Five-hundred-meter Aperture Spherical radio Telescope
\citep[FAST;][]{Nan2006-FAST,Jiang2020-FAST}
for the remaining 113 massive red spirals from \citet{Guo2020}, 
detecting \hi\ in 75 of them. 
\citet{Parkash2019} and \citet{Sharma2023} selected 
\hi-detected galaxies from \hi\ Parkes All-Sky Survey 
Catalog \citep[HICAT;][]{2004MNRAS.350.1195M} and the 
HI-MaNGA program \citep{HIMaNGA2}, 
respectively, both using a mid-infrared color selection 
of ${\rm W2}-{\rm W3}<2.0$ to select galaxies with sSFR 
$<10^{-10.4}{\rm yr}^{-1}$. \citet{Sharma2023} further 
applied a selection of \mhi$~>10^{9.3}~$\msun. 

The high detection rates of \hi\ emission in galaxies 
with suppressed SFRs appear to challenge the well-established 
correlation of low star formation with 
low \hi\ gas mass \citep[see the review by][and references therein]{Saintoge2022}. 
Based on an analysis of the data from SDSS, xGASS and ALFALFA,
\citet{Chengpeng19} claimed that nearly all massive 
quiescent disk galaxies in the local Universe have a 
large \hi\ reservoir, similar to star-forming galaxies
of similar masses. This claim was attributed by 
\citet{Cortese2020} to the use of aperture-corrected SFR 
estimates from the MPA/JHU SDSS catalog \citep{Brinchmann2004}, 
which do not provide a fair representation of the global 
SFR of \hi-rich galaxies with extended star-forming discs.
In a recent study, \citet{Li-Ho-2023} made careful 
image analysis and obtained reliable SFR estimates
for a sample of nearby galaxies, and they confirmed that 
the MPA/JHU SDSS catalog indeed significantly overestimates 
the SFRs for low-SFR galaxies. For massive red spirals, 
\citet{Zhou2021} and \citet{Wang2022-FAST} pointed out that
many of the galaxies selected by optical color $u-r$ in 
\citet{Guo2020} are actually green or even blue in \nuvr, 
the NUV-to-optical color index. 

In this work, we attempt to identify a sample of ``red but 
\hi-rich'' (RR) galaxies, in a way that differs from 
previous studies. First, we make use of three existing 
\hi\ surveys/catalogs to maximize our sample size. These 
include xGASS, ALFALFA and HI-MaNGA. Second, we use 
\nuvr\ as a reliable indicator
of the global star formation status of a galaxy, and 
we use the commonly adopted color criterion of \nuvr$~>5$ 
to select truly red, thus fully-quenched galaxies. 
As we will show, this color index works better than both 
the optical color $u-r$ and the mid-infrared color ${\rm W2}-{\rm W3}$.
Third, we use a {\em complete} (volume-limited) sample of 
galaxies selected from the SDSS as a reference, and we 
identify our RR galaxies as outliers located outside the 
95\% contour of the SDSS galaxy distribution in the 
\hi\ mass fraction versus \nuvr space. For each of 
the SDSS galaxies, we have estimated an \hi\ mass based on 
Bayesian inferences of a number of optical properties, 
a technique developed in our previous work \citep{XiaoLi}. 
With the help of the SDSS sample, we are able to quantify 
the existence and rareness of the RR galaxies as a population. 
Finally, we use SFRs from both the MPA/JHU SDSS catalog 
and the GSWLC \citep{GSWLC2}. The latter applied SED 
fitting to photometry from UV to IR, expected to provide
more reliable estimates particularly for low-SFR galaxies
\citep[e.g.][]{Li-Ho-2023}. As we will show, the 
RR sample selected this way is unique compared to all the 
previous similar samples. We will compare the optical properties 
and local environment of the RR galaxies with control samples 
of \hi-normal galaxies of similar mass and color. 

The paper is organized as follows. In \autoref{sec:data}
we introduce the data used in this analysis. We present our 
results in \autoref{sec:results} and discussion in 
\autoref{sec:discussion}. We summarize our main results in 
\autoref{sec:summary}. Throughout this paper, we assume a 
$\Lambda$CDM cosmology with $\rm \Omega_{M}=0.3$, 
$\rm \Omega_{\Lambda}=0.7$, and $h=0.7$.
We denote $\log_{10}$ as $\log$ for simplicity.

\section{Data and sample selection} \label{sec:data}

\subsection{H{\sc i} galaxy surveys}

We use  \hi\ observations of nearby galaxies 
from three surveys: ALFALFA, xGASS, and the HI-MaNGA program.

ALFALFA is a blind survey of \hi\ 21 cm emission in galaxies 
over $7000$~deg$^2$ of the sky and up to redshift $z\sim 0.06$.
Two sky areas are covered by ALFALFA: one in the northern Galactic 
hemisphere and one in the southern Galactic hemisphere. 
The final ALFALFA catalog, the $\alpha.100$ catalog 
contains $\sim 31,500$ extragalactic \hi\ line sources out to $z\sim 0.06$
\citep{2018ApJ...861...49H}. For each source, a category code 
is assigned according to data quality, with ``code 1'' for sources 
of the highest data quality and ``code 2'' for sources with lower data
quality but a high likelihood to be true. There are 25,434 sources 
of ``code 1'' and 6,068 sources of ``code 2''.
In this work we use both categories.

The xGASS was accomplished also with the Arecibo telescope
for targeted galaxies that are randomly selected from a parent sample of 
galaxies located in the overlapping region among SDSS data release 6 
\citep[][]{Adelman_McCarthy_2008}, GALEX  
Medium Imaging Survey \citep[][]{Martin_2005},
and the ALFALFA survey footprint. The whole program was 
separated into two successive phases resulting in two complementary 
surveys: GASS and xGASS. The GASS obtained \hi\ observations 
for a sample of galaxies with redshift $0.025<z<0.05$ and a flat 
stellar mass distribution in the range 
$10^{10}$\msun$<$\mstar$<10^{11.5}$\msun, 
down to a uniform limit of $\sim1.5$\% in \hi-to-stellar mass ratio.
The xGASS extends the stellar mass limit of the survey down to 
\mstar$\sim10^9$\msun, by further observing a sample of 
galaxies with $10^{9}$\msun$<$\mstar$<10^{10.2}$\msun\
in the redshift range $0.01<z<0.02$. In this work we use the 
xGASS representative sample constructed 
by \cite{xGASS}, including 1,179 galaxies from GASS and xGASS.

The HI-MaNGA is an \hi\ followup program for the SDSS-IV 
MaNGA survey \citep{Bundy_2015}. The MaNGA galaxies with 
$cz<15000\ \rm km\ s^{-1}$ and located outside the ALFALFA 
footprint are observed with the Green Bank telescope, down to 
a rms noise comparable to the ALFALFA survey (around 1.5mJy 
at a velocity resolution of 10 $\rm km\ s^{-1}$). 
The final \hi-MaNGA catalog contains 6,358 MaNGA galaxies. 

\subsection{The SDSS galaxy sample}

To compare our \hi-selected samples with the general galaxy 
population, we have selected a volume-limited sample of 
galaxies out of the New York University Value-Added Galaxy Catalog
(NYU-VAGC), constructed by \citet{Blanton_2005} from the 
SDSS spectroscopic galaxy sample. This sample includes 
69,296 galaxies with stellar masses \mstar$>10^9$\msun\ 
and spectroscopically-measured redshifts $0.02<z<0.05$.

\subsection{Galaxy properties}

For each galaxy in the \hi\ surveys and the SDSS sample 
we have collected or measured the following properties: 
\begin{itemize}
    \item \mstar: stellar mass from the NASA-Sloan 
    Atlas\footnote{http://nsatlas.org} \citep[NSA;][]{2011AJ....142...31B},
    estimated by \citet{2007AJ....133..734B} by performing a 
    spectral energy distribution (SED) fitting to SDSS photometry 
    assuming a Chabrier stellar initial mass function\citep{Chabrier2003}.
    \item \nuvr: color index defined by NUV and $r$-band 
    absolute magnitudes provided in NSA, both 
    $K$-corrected and measured with an ellipitcal Petrosian model,
    but based on the GALEX and SDSS images, respectively.
    \item $\sigma_{{\rm NUV}-r}$: error of \nuvr, calculated through 
    error propagation based on the inverse variance of the NUV and 
    $r$ band absolute magnitudes in the NSA.
    \item $R_{50}$ and $R_{90}$: the radii containing, respectively, $50\%$ and $90\%$
    of the elliptical Petrosian flux in $r$ band, in units of kpc, taken from NSA.
    \item $R_{90}/R_{50}$: the concentration index characterizing 
    the degree of light concentration in the $r$ band.
    \item $\mu_\ast\equiv M_\ast/(2\pi R_{50}^2)$: surface stellar mass density
    in units of $\rm M_{\odot}\ kpc^{-2}$.
    \item $D_{4000}$: the spectral break at $4000$\AA\ taken from 
    the MPA/JHU SDSS catalog\footnote{https://wwwmpa.mpa-garching.mpg.de}, 
    measured from the SDSS single fiber spectroscopy following the definition 
    introduced by \citet{1999ApJ...527...54B}.
    \item $\sigma_\ast$: central stellar velocity dispersion taken from the 
    MPA/JHU SDSS catalog, in units of km/s. 
    \item SFR: total star formation rate in units of $\rm M_{\odot}\ yr^{-1}$.
    We use two SFR estimates: one from the GSWLC catalog derived by 
    \citet{GSWLC2} through SED fitting to the photometry ranging from UV to IR, and 
    the other from the MPA/JHU catalog derived by \citet{Brinchmann2004} based 
    on SDSS spectroscopy. The Chabrier IMF is assumed in both cases. 
    Unless otherwise stated we will use the GSWLC-based SFR.  
    \item sSFR~$\equiv {\rm SFR}/M_*$: specific star formation rate.
    The GSWLC-based SFR is used unless otherwise stated.
    \item $B/T$: bulge-to-total luminosity ratio, taken from \cite{Meert_2015}.
    \item T-type: galaxy morphological type provided by
    \cite{2018MNRAS.476.3661D}. Generally, galaxies with T-type~$<0$ 
    (T-type~$>0$) correspond to early-type (late-type) morphologies in 
    the Hubble sequence.
    \item $P_{merger}$: probability of being a merger or projected pair taken from \cite{2018MNRAS.476.3661D}, 
    based on the model trained by the Galaxy Zoo 2 catalog 
    \citep[GZ2;][]{Willett_2013}.
    \item $P_{bar}$: probability of having a galactic bar taken from \cite{2018MNRAS.476.3661D}.
    Two estimates are provided for each galaxy, $P_{\rm bar,GZ2}$ and $P_{\rm bar,N10}$,
    based on two different models trained by the GZ2 catalog and the 
    catalog of \cite{Nair_2010}, respectively. We use both estimates in our work.
    \item $(g-r)_{R_{50}}$ and $(g-r)_{0.5R_{50}}$: color index $g-r$ measured 
   at $R_{50}$ and $0.5R_{50}$, based on the $g$ and $r$ band images from
   the DESI Legacy Survey\footnote{https://www.legacysurvey.org}.   
    \item $\Delta_{g-r}\equiv (g-r)_{R_{50}}-(g-r)_{0.5R_{50}}$: 
    radial gradient in $g-r$. 
\end{itemize}

In addition to the optical properties listed above, we have estimated 
the \hi\ gas mass for each galaxy using the estimator developed in 
\citet{XiaoLi}. In short, the estimator uses a linear combination of four 
photometric parameters to estimate the \hi-to-stellar mass 
ratio (\mhi/\mstar) of galaxies, with coefficients of the parameters 
constrained using Bayesian inferences of real \hi\ data from the xGASS.
Extensive tests show that the estimator can provide unbiased \hi\ 
mass for optically-selected samples like SDSS. The reader is referred 
to \citet{XiaoLi} for more details about the tests and applications of the 
estimator. The \hi\ mass distribution of the SDSS volume-limited sample 
provides a {\em predicted} reference, to be compared below with  
samples selected from the \hi\ surveys. 

\begin{figure*}[ht!]
	\centering
	\includegraphics[width=0.9\textwidth]{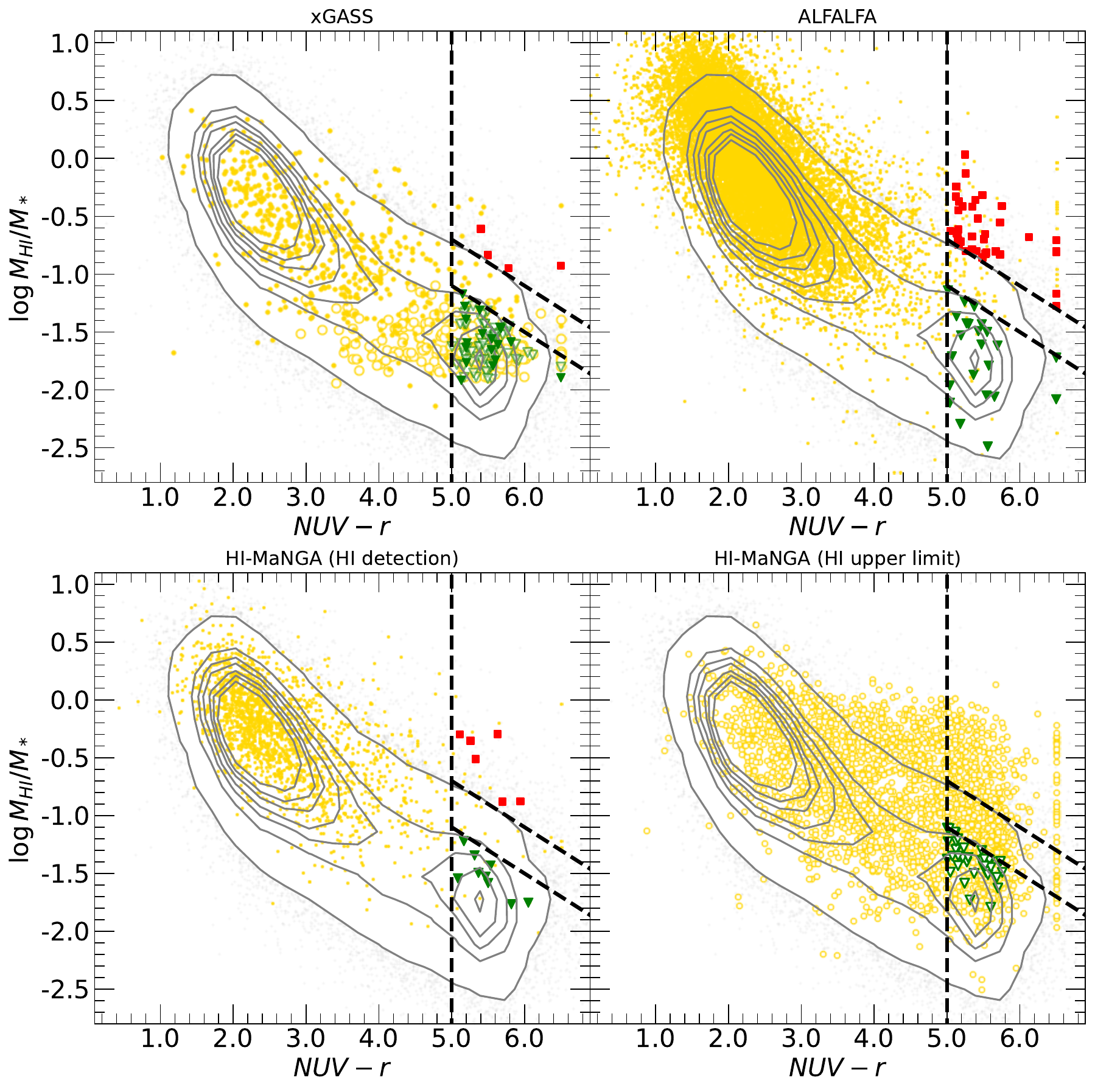}
	\caption{Sample selection. The gold solid dots represent the parent sample of all galaxies with \hi\ observation from xGASS (upper-left), ALFALFA (upper-right), and \hi-MaNGA (lower-left for detections and lower-right for upper limits), after removing potentially confused galaxies
	and galaxies with $\sigma_{NUV-r}>0.3$. Solid and open symbols represent \hi\ detections and upper limits separately. Galaxies included in the final RR and RN sample are shown as red squares and green triangles, respectively. The color is set to $\mathrm{NUV}-r=6.5$ for a few galaxies which have large uncertainties in NUV (see text in  \autoref{sec:sample_selection} for details). The grey contours represent the SDSS volume-limited sample, of which the \hi\ mass of each galaxy is predicted by the estimator developed by \citet{XiaoLi}. The outermost contour includes $95\%$ of the total sample.}
	\label{fig:sample_selection}
\end{figure*}

\renewcommand{\arraystretch}{0.8}
\begin{deluxetable*}{cccccccccccc}[ht!]
	\label{tbl:galaxylist}
    \tablenum{1}
    \tablecaption{Basic properties of the RR sample: source id, right ascension, declination, redshift, stellar mass from the NSA, $\nuvr$ color index (those assigned a fixed $\nuvr$ are shown as $\geq 6.50$), Petrosian 50\% and 90\% radii in $r$ band, SFR taken from GSWLC and MPA/JHU (those not matched to these catalogs are denoted as --), \hi\ mass, source of \hi\ mass (0:ALFALFA, 1:xGASS, 2:\hi-MaNGA)\label{tab:rr_info}}
    \tablewidth{0pt}
    \tablehead{
        \colhead{id} & \colhead{ra} & \colhead{dec} & \colhead{z} & \colhead{$\lgmstar$} & \colhead{$\nuvr$} & \colhead{$\rm \log_{10}R_{50}$} & \colhead{$\rm \log_{10}R_{90}$} & \colhead{$\rm \log_{10} SFR_{GSWLC}$} & \colhead{$\rm \log_{10} SFR_{MPA}$} & \colhead{$\lgmhi$} & \colhead{\hi\ src} \\
        \colhead{} & \colhead{} & \colhead{} & \colhead{[$\msun$]} & \colhead{[mag]} & \colhead{[kpc]} & \colhead{[kpc]} & \colhead{[$\rm \msun \ yr^{-1}$]} & \colhead{[$\rm \msun \ yr^{-1}$]} & \colhead{[$\msun$]} & \colhead{}
    }
    \startdata
    0 & 28.6710 & -0.1434 & 0.0185 & 10.68 & 5.33 & 0.54 & 1.06 & -1.37 & -1.54 & 10.17 & 2 \\
    1 & 19.4449 & 13.3234 & 0.0479 & 10.25 & 5.26 & 0.33 & 0.86 & -1.87 & -1.22 & 10.12 & 0 \\
    2 & 128.1644 & 53.2332 & 0.0430 & 10.30 & 5.11 & 0.39 & 0.88 & -1.99 & -0.98 & 10.00 & 2 \\
    3 & 52.3316 & -6.4779 & 0.0420 & 10.20 & 5.26 & 0.23 & 0.67 & -1.02 & -1.17 & 9.85 & 2 \\
    4 & 149.4660 & 1.7554 & 0.0320 & 10.54 & 5.70 & 0.39 & 0.83 & -2.05 & -1.23 & 9.66 & 2 \\
    5 & 179.6394 & 1.3700 & 0.0485 & 10.32 & 5.73 & 0.26 & 0.71 & -1.86 & -1.39 & 9.77 & 0 \\
    6 & 142.0508 & 3.4084 & 0.0116 & 10.31 & 5.53 & 0.11 & 0.62 & -2.25 & -1.66 & 9.48 & 0 \\
    7 & 153.3518 & 5.0255 & 0.0464 & 10.05 & 5.40 & 0.14 & 0.55 & -2.05 & -1.38 & 9.44 & 1 \\
    8 & 221.8548 & 3.4411 & 0.0274 & 10.42 & 5.39 & 0.29 & 0.82 & -1.37 & -1.29 & 10.06 & 0 \\
    9 & 333.9182 & 13.8821 & 0.0246 & 10.09 & 5.15 & 0.30 & 0.71 & -0.65 & -1.13 & 9.38 & 0 \\
    10 & 340.1470 & 13.9773 & 0.0367 & 10.24 & 5.35 & 0.29 & 0.78 & -0.78 & -1.31 & 9.82 & 0 \\
    11 & 198.3802 & 5.9155 & 0.0486 & 10.98 & 5.43 & 0.73 & 1.15 & -1.32 & -0.95 & 10.46 & 0 \\
    12 & 163.0032 & 50.8430 & 0.0260 & 10.20 & 5.94 & 0.30 & 0.73 & -2.58 & -1.43 & 9.32 & 2 \\
    13 & 159.7155 & 5.6701 & 0.0283 & 10.26 & 5.63 & 0.33 & 0.77 & -2.50 & -1.33 & 9.96 & 2 \\
    14 & 139.6306 & 6.8732 & 0.0393 & 10.38 & 5.78 & 0.26 & 0.75 & -0.68 & -1.29 & 9.44 & 1 \\
    15 & 242.0909 & 26.5281 & 0.0540 & 10.56 & 6.50 & 0.33 & 0.83 & -1.10 & -1.05 & 9.76 & 0 \\
    16 & 249.3657 & 26.2047 & 0.0522 & 10.81 & 6.50 & 0.81 & 1.25 & -1.21 & -1.24 & 10.10 & 0 \\
    17 & 215.6282 & 11.3049 & 0.0164 & 10.36 & 5.34 & 0.10 & 0.61 & -1.49 & -1.80 & 9.57 & 0 \\
    18 & 223.2200 & 12.3523 & 0.0299 & 10.50 & 5.73 & 0.32 & 0.82 & -1.94 & -1.46 & 9.67 & 0 \\
    19 & 155.5138 & 13.7695 & 0.0186 & 10.18 & 5.14 & 0.36 & 0.88 & -2.92 & -1.42 & 9.53 & 0 \\
    20 & 226.2269 & 7.2125 & 0.0560 & 10.09 & 5.25 & 0.25 & 0.67 & -1.52 & -1.45 & 10.12 & 0 \\
    21 & 216.0679 & 33.6126 & 0.0223 & 10.38 & 5.26 & 0.24 & 0.68 & -1.95 & -1.62 & 9.58 & 0 \\
    22 & 149.7208 & 31.6203 & 0.0214 & 10.47 & 5.49 & 0.37 & 0.88 & -2.19 & -1.36 & 9.61 & 0 \\
    23 & 178.1931 & 32.3099 & 0.0317 & 10.44 & 5.16 & 0.27 & 0.78 & -1.93 & -0.98 & 9.82 & 0 \\
    24 & 205.5474 & 29.8956 & 0.0272 & 10.22 & 5.16 & 0.02 & 0.47 & -1.56 & -1.59 & 9.77 & 0 \\
    25 & 208.8057 & 25.2164 & 0.0360 & 10.22 & 5.49 & 0.31 & 0.80 & -1.42 & -1.30 & 9.90 & 0 \\
    26 & 227.1210 & 21.9477 & 0.0206 & 10.56 & 5.34 & 0.45 & 0.94 & -2.04 & -1.60 & 9.89 & 0 \\
    27 & 246.7624 & 16.3822 & 0.0154 & 10.46 & 6.50 & 0.52 & 0.97 & -1.48 & -1.25 & 9.65 & 0 \\
    28 & 142.5893 & 25.2524 & 0.0587 & 10.95 & 5.41 & 0.71 & 1.24 & -0.77 & -1.23 & 10.15 & 0 \\
    29 & 162.3703 & 26.0396 & 0.0217 & 10.36 & 5.51 & 0.16 & 0.66 & -2.16 & -1.45 & 9.66 & 0 \\
    30 & 155.4069 & 21.6873 & 0.0392 & 10.53 & 5.16 & 0.36 & 0.83 & -0.51 & -1.26 & 9.78 & 0 \\
    31 & 120.0874 & 11.3194 & 0.0149 & 10.25 & 6.13 & 0.27 & 0.68 & -1.40 & -1.90 & 9.57 & 0 \\
    32 & 132.6887 & 11.8108 & 0.0297 & 10.41 & 5.50 & 0.20 & 0.66 & -- & -1.36 & 9.58 & 1 \\
    33 & 241.2004 & 14.1289 & 0.0338 & 10.87 & 5.12 & 0.42 & 0.91 & -1.71 & -1.09 & 10.54 & 0 \\
    34 & 240.8469 & 14.5245 & 0.0357 & 10.52 & 5.19 & 0.29 & 0.77 & -1.40 & -1.28 & 9.80 & 0 \\
    35 & 150.2089 & 13.7370 & 0.0325 & 10.20 & 5.16 & 0.09 & 0.60 & -1.90 & -1.51 & 9.83 & 0 \\
    36 & 171.2433 & 7.6274 & 0.0415 & 10.75 & 5.05 & 0.48 & 1.01 & -- & -- & 10.12 & 0 \\
    37 & 16.5919 & 6.2825 & 0.0352 & 10.87 & 5.13 & 0.48 & 0.98 & -- & -- & 10.63 & 0 \\
    38 & 23.6161 & 19.8206 & 0.0357 & 10.87 & 5.14 & 0.49 & 0.96 & -- & -- & 10.22 & 0 \\
    39 & 45.1559 & 35.1691 & 0.0169 & 10.77 & 6.50 & 0.58 & 0.98 & -- & -- & 9.50 & 0 \\
    40 & 359.4070 & 18.1909 & 0.0260 & 10.53 & 5.54 & 0.24 & 0.71 & -- & -- & 9.72 & 0 \\
    41 & 1.4582 & 17.3534 & 0.0181 & 10.17 & 5.21 & 0.04 & 0.55 & -- & -- & 9.76 & 0 \\
    42 & 24.0705 & 23.2281 & 0.0343 & 10.93 & 5.67 & 0.59 & 1.11 & -- & -- & 10.13 & 0 \\
    43 & 329.8712 & 15.2784 & 0.0252 & 10.50 & 5.53 & 0.23 & 0.73 & -- & -- & 9.85 & 0 \\
    44 & 37.0733 & 26.3125 & 0.0174 & 10.63 & 5.75 & 0.26 & 0.80 & -- & -- & 10.22 & 0 \\
    45 & 344.9648 & 26.0331 & 0.0332 & 11.15 & 6.50 & 0.93 & 1.34 & -- & -- & 9.98 & 0 \\
    46 & 207.9975 & 13.9675 & 0.0366 & 10.89 & 6.50 & 0.83 & 1.28 & -0.21 & 0.04 & 9.96 & 1
    \enddata
\end{deluxetable*}

\subsection{Selection of red \hi-rich galaxies} \label{sec:sample_selection}

\autoref{fig:sample_selection} shows 
$\log_{10}$\mhi/\mstar\ versus \nuvr, for the three \hi\ surveys and 
the SDSS volume-limited galaxy sample. The SDSS sample is 
plotted as grey contours and repeated in every panel.
Galaxies in different \hi\ surveys are shown as colored symbols 
in different panels. In each panel, \hi\ detections are plotted 
as solid dots, while upper limits are plotted as open circles. 
For clarity the \hi-MaNGA sample is divided into two panels, one for 
\hi\ detections and one for upper limits. The anti-correlation 
between \hi\ mass fraction and color, and the color bimodality
can be clearly seen in the SDSS sample. All the current 
\hi\ surveys are biased for relatively \hi-rich and blue galaxies 
due to limited survey depths. Although the smallest in sample size,
the xGASS is the deepest and thus most complete among all 
the surveys considered here. However, even the xGASS lacks all the 
red galaxies with \mhi/\mstar$\lesssim 1.5$\%. Interestingly, 
the shallower ALFALFA survey has detected \hi\  
for a (small) number of low-$z$ galaxies with red colors (\nuvr$>5$) and 
low gas fraction because of their relatively high \hi\ masses, 
covering nearly the same range of \mhi/\mstar\ as predicted for the red galaxy 
population in the SDSS.

\begin{figure*} [htbp]
	\centering
	\includegraphics[width=\textwidth]{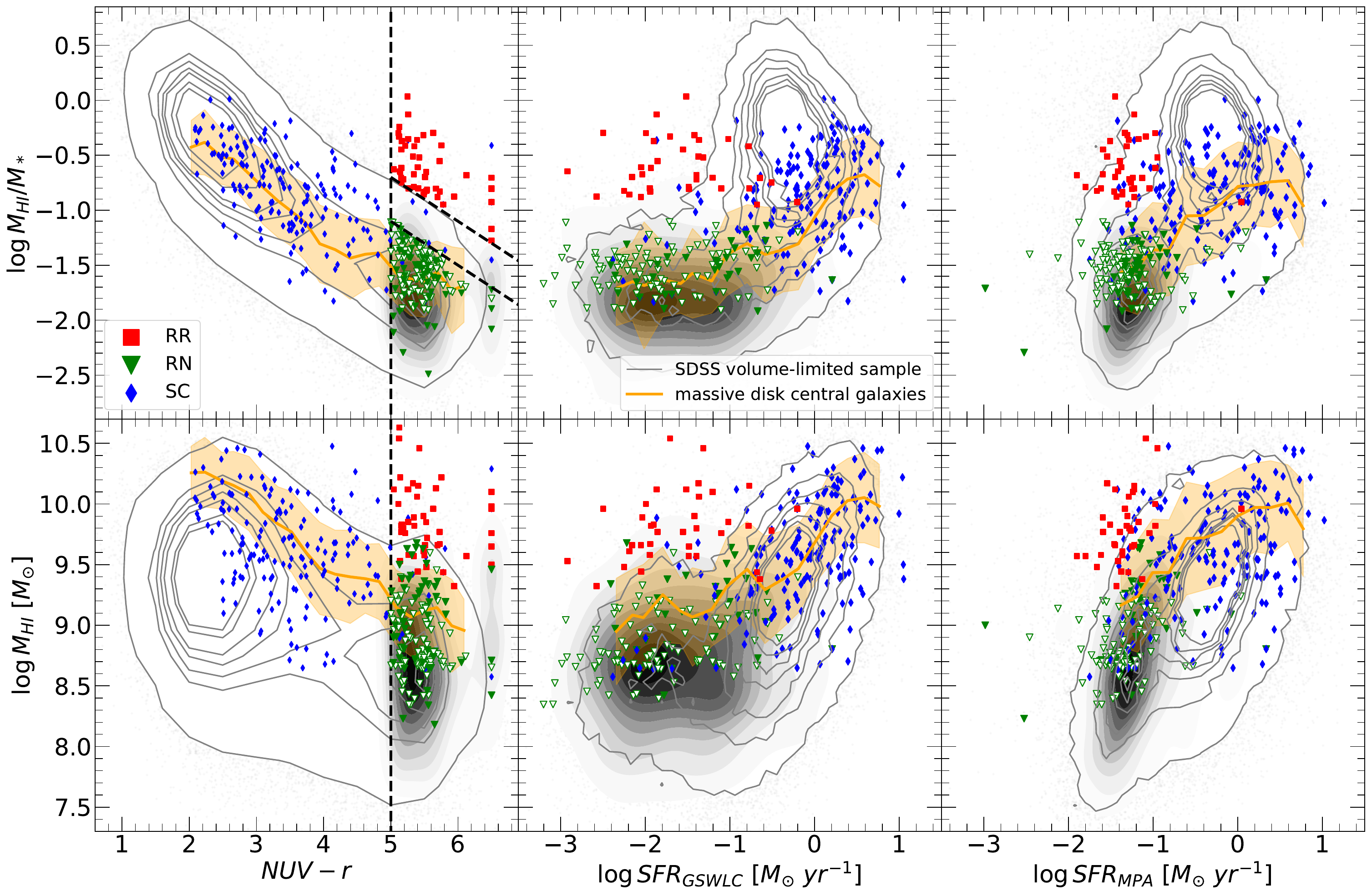}
	\caption{Diagrams of $\mhi/\mstar$ (the top row) and $\mhi$ (the bottom row), versus $NUV-r$ (the left column) and SFRs (with SFRs from GSWLC in the middle column, and SFRs from the MPA/JHU catalog in the right column). The RR, RN, and SC galaxies are shown as red 
	squares, green triangles, and blue diamonds, with solid and open symboles for \hi\ detections and non-detections (upper limits) separately. The color is set to $\mathrm{NUV}-r=6.5$ for a few galaxies which have large uncertainties in NUV (see text in \autoref{sec:sample_selection} for details). The grey contours represent the SDSS volume-limited sample, of which the \hi\ mass of each galaxy is predicted by the estimator developed by \citet{XiaoLi}. The outermost contour includes $95\%$ of the total sample.
	The filled grey contours represent the CSC sample. The orange solid line and the shaded region show the median and the $1-\sigma$ scatter of the \hi-to-stellar mass ratio for massive disk central galaxies.
	\label{fig:fHI_vs_SFR}}
\end{figure*}

We select red but \hi-rich (RR) galaxies from the three \hi\ surveys 
by requiring them to have substantially red colors, 
$\nuvr>5$ and $\sigma_{{\rm NUV}-r}<0.3$, 
as well as outstandingly high \hi\ masses, 
$\log_{10}($\mhi/\mstar$)+0.4\times({\rm NUV}-r) - 1.3 > 0$.
The two criteria are indicated by the vertical and upper-right 
dashed lines in \autoref{fig:sample_selection}. The latter criterion 
is manually determined, and it roughly matches the 95\% 
percentile contour of the SDSS galaxy distribution. 
We consider only \hi\ detections when identifying RR galaxies,
to make sure their \hi\ mass fractions meet our requirement.
For comparison, we have selected a sample of red and \hi-normal 
(RN) galaxies with the same criteria in color and color error
as adopted for RR, but requiring the \hi-to-stellar mass ratio to fall below 
the 75\% percentile contour, or specifically
$\log_{10}$\mhi/\mstar$+0.4\times({\rm NUV}-r) - 0.9 < 0$,
as indicated by the lower dashed line plotted to the right of each panel.
We consider both \hi\ detections and upper limits when selecting
RN galaxies. This is because a galaxy must fall below the 
lower dashed line if its \hi\ upper limit is already 
below this line. There are a small number of galaxies with unusually 
red colors (\nuvr$~\gg~$6.5) due to their extremely large values 
of NUV magnitude with large uncertainties. 
We have visually examined their NUV images and found them 
to be truly faint in NUV. In order to include them in our 
analysis, we assign them a fixed color index of \nuvr~$=6.5$,
the reddest color that can be reached by a simple stellar 
population (SSP) in current stellar population models
\citep[e.g.][]{BC03}. We apply the same criteria as above
to select these galaxies into the RR or RN sample. 
In addition, we require all the selected RR and RN 
galaxies to not be confused with neighboring galaxies located in 
the same beam. We identify galaxies free of confusion
by requiring {\qcr HIconf\_flag=0} for those from xGASS, 
and {\qcr conf\_prob<0.1} for those from \hi-MaNGA.
For those from ALFALFA, we require them to have 
no neighboring galaxies within an angular radius of 3.5$^\prime$ 
and a radial velocity separation of $600\rm\ km s^{-1}$.
Finally, we visually examine the DESI image of all 
the selected galaxies, and exclude the edge-on galaxies 
with obvious dust lanes which could be intrinsically blue 
but reddened by dust attenuation. 

With these restrictions we end up with 47 RR galaxies 
(37 from ALFALFA, 4 from xGASS and 6 from \hi-MaNGA),
as well as 573 RN galaxies (43 from ALFALFA, 
237 from xGASS and 293 from \hi-MaNGA).
The basic properties of the RR sample are listed in \autoref{tab:rr_info}.
In order to take care of the mass dependence when 
comparing the RR and RN samples, we further trim the 
RN sample so that its \mstar\ distribution is the same 
as the RR sample. This yields a final sample of 
154 RN galaxies, with 23 from ALFALFA, 80 from xGASS 
and 51 from \hi-MaNGA. The RR and RN galaxies are highlighted 
in red and green color in \autoref{fig:sample_selection}, respectively.  
Note that the RN sample is dominated by upper limits. 

\begin{figure*} [t!]
	\centering
	\includegraphics[width=\textwidth]{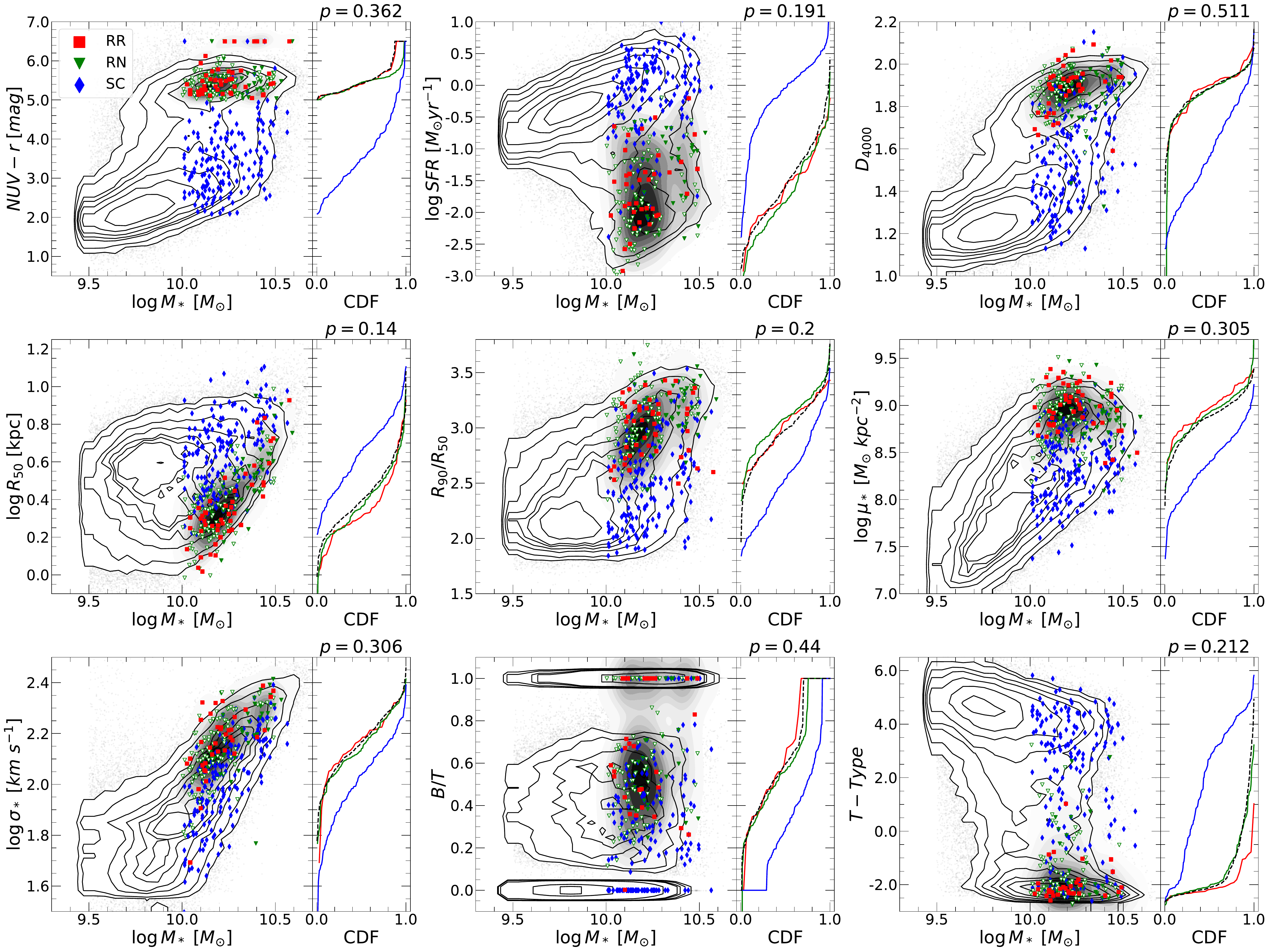}
	\caption{Galaxy properties as a function of stellar mass. The grey contours represent the SDSS volume-limited sample, of which the HI mass of each galaxy is predicted by the HI estimator \citep{XiaoLi}. In this figure, some galaxies have $NUV-r=6.5$, this is not the observed value but set artificially (see \autoref{sec:sample_selection} for details. Solid/open markers represent \hi\ detections/non-detections(upper limits).
    The red squares represent RR galaxies. 
    The green downward triangles represent RN galaxies.
    The blue diamonds represent SC galaxies.
    The colored lines denote the
    normalized cumulative distribution of the y-axis galaxy properties (red for RR, green for RN, blue for SC, 
    and black dashed for CSC). 
    The grey filled contours represent the CSC sample. The p-value of the two-sample KS test conducted between the RR and the RN sample is indicated above each side panel.}
	\label{fig:diagrams}
\end{figure*}

\begin{figure*} [htbp!]
	\centering
	\includegraphics[width=\textwidth]{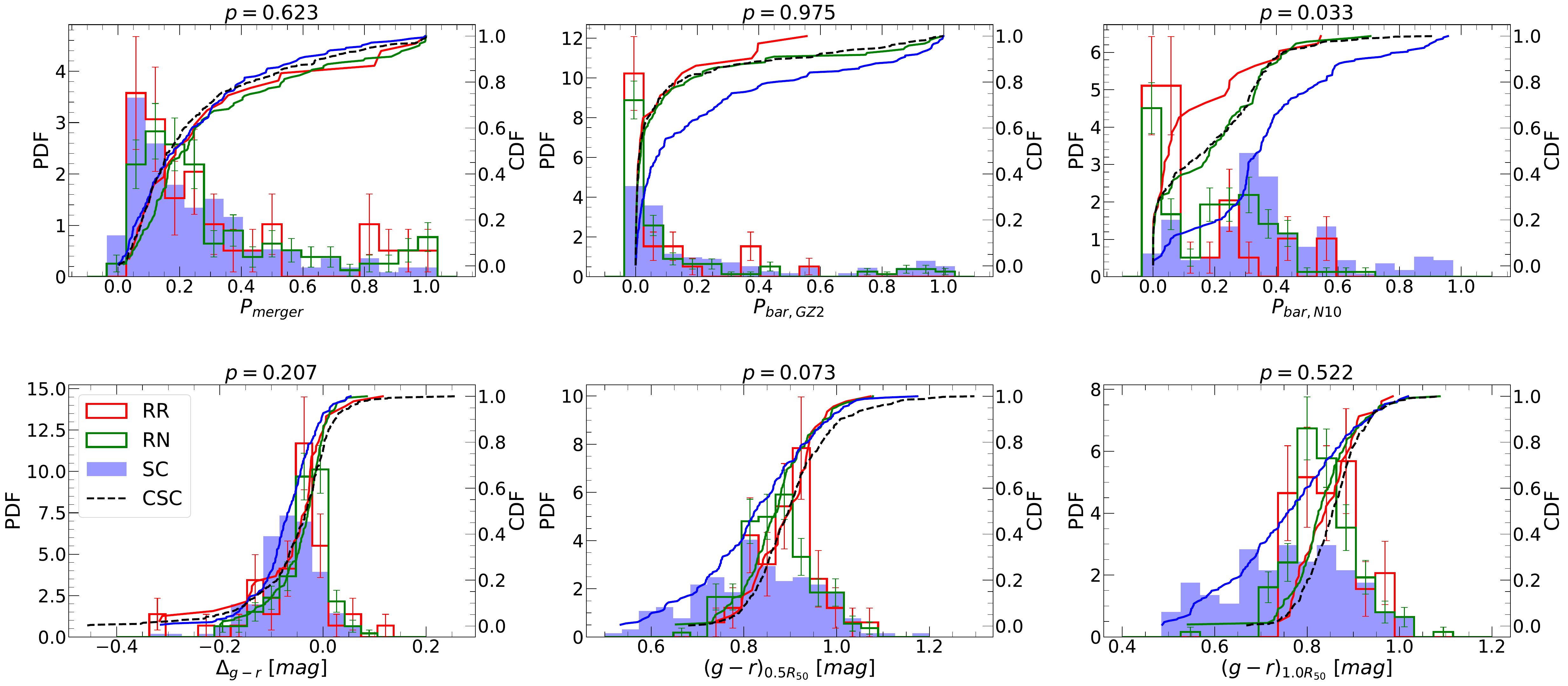}
	\caption{Probability distribution of galaxy properties. From top left to bottom right, the quantities in each panel are
    probability of being a merger or projected pair ($P_{merger}$), the probability of having
    bar signatures ($P_{bar,GZ2}$ and $P_{bar,N10}$), 
    $g-r$ color gradient, $g-r$ at $0.5R_{50}$,
    and $g-r$ at $R_{50}$. The colored solid lines show the 
    normalized cumulative distribution of the histograms with the same color. The p-value of the two-sample KS test conducted between the RR and the RN sample is indicated above each panel.}
	\label{fig:histograms}
\end{figure*}

The upper-left panel of \autoref{fig:fHI_vs_SFR}
displays the \mhi/\mstar\ versus \nuvr\ relation for the 
final samples of RR and RN galaxies. 
In addition to these two samples, we have selected a 
stellar mass-controlled (SC) sample of 177 galaxies from 
the xGASS, which has the same \mstar\ distribution 
as the RR and RN samples but with no restrictions on 
the color and the \hi\ mass fraction. The galaxies in the SC sample
are plotted as blue diamonds. The lower-left panel 
plots the \hi\ mass instead of the \hi-to-stellar mass 
ratio as a function of \nuvr\ for the same samples. 
In the middle and right panels we plot \mhi/\mstar\
and \mhi\ as functions of SFR, using the SFR estimates 
from GSWLC and the MPA/JHU catalog, respectively.

The RR and RN galaxies are roughly separated at 
\mhi$\sim 10^{9.5}$\msun, and both samples are dominated 
by galaxies with low SFRs ($\log_{10}$SFR$\lesssim-1$)
consistent with their red colors, for SFRs given by both 
GSWLC and MPA/JHU. When the GSWLC SFRs are used
and at the low-SFR end ($\log_{10}$SFR$\lesssim-1$), 
the RN and SC samples appear to follow roughly  
the overall trend of the SDSS sample, where both the 
\hi-to-stellar mass ratio and the total \hi\ mass 
depend only weakly on the SFR, roughly
at $\log_{10}$(\mhi/\mstar)$\sim-1.8$ and 
$\log_{10}$(\mhi/\msun)$\sim8.7$, respectively. At higher SFRs, 
the SC sample follows the \mhi-SFR relation of the SDSS 
sample, but is biased towards relatively low $\log_{10}$(\mhi/\mstar). 
This should be produced by the relatively high stellar 
masses of the SC sample at which the average SFR is 
reduced by galaxies from the quiescent sequence 
(see \autoref{fig:diagrams} below). 
When the MPA/JHU SFRs are used and at the low-SFR end, 
the SDSS sample presents a clump-like distribution 
in the \mhi/\mstar\ versus SFR relation and a positive 
correlation between \mhi\ and SFR, with a more limited 
SFR range and larger scatter in \mhi/\mstar\ and \mhi
than when the GSWLC SFRs are adopted. 
The distributions of the RR, RN and SC samples relative 
to the SDSS sample remain similar, however, regardless 
of the adopted SFR estimate. 

One may worry that the RN sample is not substantially 
	representative of the general population of red galaxies, as it 
	is selected from the \hi\ surveys but not directly from the parent 
	SDSS sample. To address this possibility, we use the SDSS 
	sample to construct another control sample (referred to as ``CSC''), 
	which consists of 423 galaxies and is matched with the RR sample 
	in both \mstar\ and \nuvr. The distributions of the CSC sample 
	are shown as filled grey contours in \autoref{fig:fHI_vs_SFR}. 
	As expected, the CSC sample well represents the underlying red 
	population of the SDSS sample.

The majority of the RR galaxies are located beyond the 
95\% contour of the SDSS sample, suggesting that red 
galaxies with large amounts of \hi\ gas are rather rare. 
In other words, red/quenched galaxies are mostly \hi-poor, 
with only a small fraction having unusually large amounts 
of \hi\ gas. In the next section we will compare the RR and
RN samples in detail, aiming to figure out the origin 
of the high \hi\ gas content of RR galaxies.

\section{Results} \label{sec:results}

\subsection[short]{Optical properties}

In \autoref{fig:diagrams} we examine the optical properties 
for the RR and RN samples, and compare with the SC sample
and the SDSS volume-limited sample. We consider a variety of 
galaxy properties, each plotted in one panel as a function 
of $\log_{10}$\mstar.
In each panel, the RR, RN and SC galaxies are plotted 
as red squares, green triangles, and blue diamonds, respectively. 
The distribution of the SDSS 
sample is plotted as the background contours. 
The CSC sample is shown as the grey filled contours.
A side panel is added to the right of each panel, showing the cumulative 
distribution of the corresponding property for the RR, 
RN, SC and CSC samples, respectively. 
Overall, the well-known galaxy bimodality can be clearly 
seen from the SDSS sample in many properties, e.g. \nuvr, 
SFR, $D_{4000}$, $R_{90}/R_{50}$, $\mu_\ast$ and the T-type.
In all panels the RN sample shows similar distributions 
	to the CSC sample, demonstrating that the RN sample is a
	representative subsample of the red galaxy population of 
	their mass.
The RR galaxies are limited to relatively high masses, with 
\mstar$\ga 10^{10}$\msun, a result that can be understood 
from their red colors. As can be seen from the color-mass 
diagram (the top-left panel), the general 
population of galaxies at \nuvr$>5$ from the SDSS sample is predominantly 
massive.

We find that the RR and RN galaxies have similar distributions 
in all the properties considered. Both types of galaxies 
fall in the sequences of red (by selection) and quiescent 
galaxies with relatively low SFRs and old stellar populations 
as indicated by their high $D_{4000}$. The two types of 
galaxies are also similar in optical size ($R_{50}$) and 
structural properties ($R_{90}/R_{50}$ and $\mu_\ast$) 
at any given mass, with smaller sizes, more centrally 
concentrated light distributions and higher stellar mass 
densities in comparison to the general galaxy population. 
In addition, they exhibit similarly high $\sigma_\ast$, 
high $B/T$ and low T-type, all indicative of early-type 
morphologies.  

In \autoref{fig:histograms} we further examine the merger 
probability ($P_{\rm merger}$), the probabilities of hosting 
a galactic bar ($P_{\rm bar,GZ2}$ and $P_{\rm bar,N10}$),
the $(g-r)$ color indices measured at $R_{50}$ and $0.5R_{50}$,
and the radial gradient of the color indices. For each 
property, we show both the differential and the cumulative 
distributions for the three samples: RR, RN and SC. 
The RR and RN samples show very similar distributions
for all properties considered except $P_{\rm bar,N10}$, 
which appears to be smaller in the RR sample on average. 
The different distributions of $P_{\rm bar,GZ2}$ and 
$P_{\rm bar,N10}$ may be produced by the different 
definitions of the two parameters, which can also be seen 
from their distributions in the SC sample. Given 
	this difference, both the large $p$-value of the 
	KS test using $P_{\rm bar,GZ2}$ and the small $p$-value 
	using $P_{\rm bar,N10}$ should be taken with caution.
 We note that,
however, nearly all the galaxies in both samples have 
$P_{\rm bar}<0.5$ indicative of no/weak bars, a result 
that is true for both $P_{\rm bar,GZ2}$ and $P_{\rm bar,N10}$.
The majority of the RR and RN galaxies have 
low merger and bar probabilities, as well as negative color 
gradients, with $P_{\rm merger}<0.5$, $P_{\rm bar, GZ2}<0.2$, 
$P_{\rm bar, N10}<0.3$ and $\Delta_{g-r}<0$ for
$\sim80\%$ of the sample galaxies. Nearly all the 
RR and RN galaxies have $(g-r)>0.7$ at both $0.5R_e$ and 
$R_e$, consistent with their globally red colors.  
When compared to the SC sample, 
both the RR and RN samples are similar in $P_{\rm merger}$ 
and $\Delta_{g-r}$, smaller in $P_{\rm bar,GZ2}$ and 
$P_{\rm bar,N10}$, and larger in $(g-r)$ at both $R_{50}$ 
and $0.5R_{50}$. 

Results as seen from \autoref{fig:diagrams} and 
\autoref{fig:histograms} strongly suggest that, when compared 
to the RN galaxies of similar stellar masses, the RR 
galaxies have nothing special in any of the properties
we have examined, including both the properties related 
to stellar populations and those related to galaxy structure, 
morphology and mergers. In fact, for each property we 
have performed a Kolmogorov-Smirnov (K-S) test 
for the RR and RN samples. The resulting $p$ values are 
indicated above the panels in both figures, which 
confirm that the two samples are statistically drawn 
from the same parent sample.

\begin{figure*}
    \centering
    \includegraphics[width=0.45\textwidth]{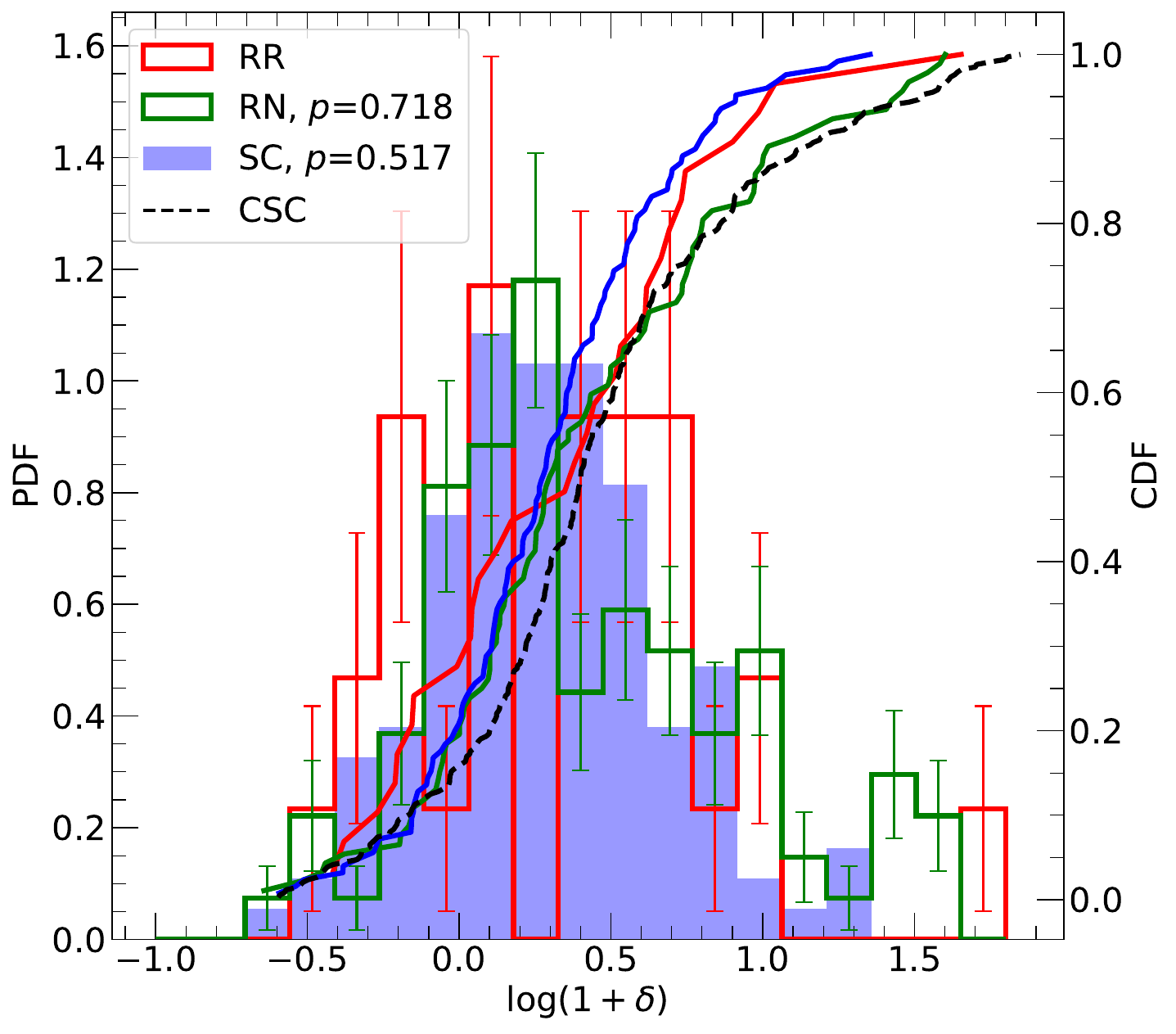}
    \includegraphics[width=0.45\textwidth]{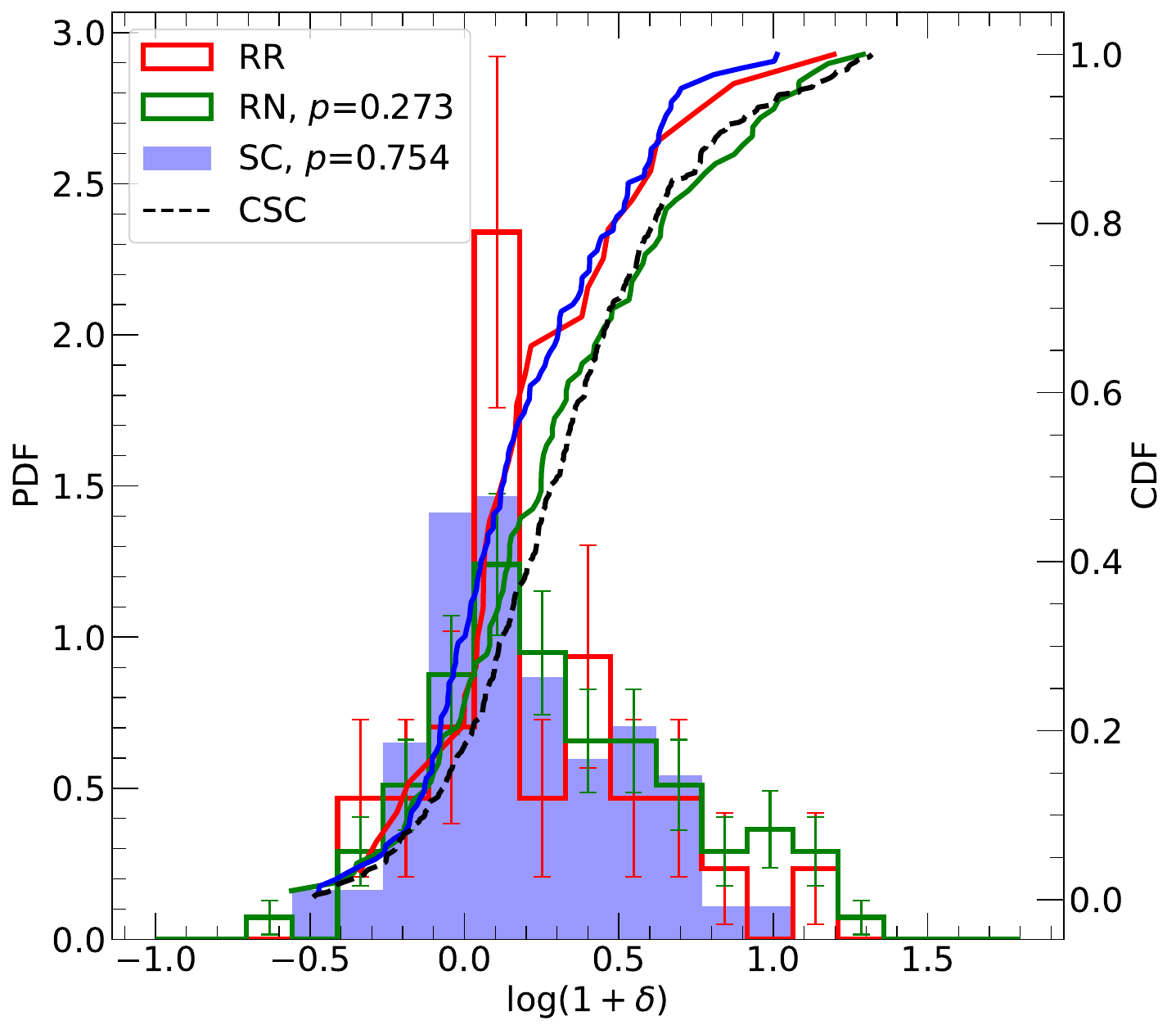}
    \caption{Histograms and cumulative distributions of $\log_{10}(1+\delta)$, 
    where $\delta$ is the local overdensity estimated at scales of 2 Mpc (left)
    and 4 Mpc (right) by \citet{Wang2016}. Results for the RR, RN, SC and CSC
    samples are plotted with different colors as indicated. 
    For clarity,  the histogram is not shown for the CSC sample.
    The $p$ values from K-S tests are indicated by comparing the RR sample 
    with the RN and SC samples.}
    \label{fig:overdensity}
\end{figure*}

\subsection[short]{Environments} \label{subsec:group}

The cold gas content of galaxies may be effectively reduced by 
environmental effects occurring in dense regions such as tidal 
stripping and ram-pressure stripping. These effects are expected 
to be more efficient for satellite galaxies than for central galaxies.
To examine the central/satellite fractions of our galaxies and 
their environments at different scales, we use the SDSS 
galaxy group catalog of \citet{2007ApJ...671..153Y} and the 
local environment density inferred from the density field reconstructed 
by \citet{Wang2016}, and we measure the projected two-point 
correlations and neighbor counts for our galaxy samples using 
the SDSS spectroscopic and photometric samples. 

First of all, we cross match the RR, RN and SC samples with the 
group catalog constructed by \cite{2007ApJ...671..153Y}. A small 
number of our galaxies are not included in the group catalog, 
including 10, 16, and 3 galaxies from the RR, RN and SC samples, 
respectively.
For these galaxies, we use the NSA to examine their neighboring 
galaxies, and we find that most of them (10/10, 14/16, 3/3 for the 
RR, RN and SC samples, respectively) are either the most massive or the only 
galaxy within a projected radius of 1 Mpc. We thus classify these 
``locally dominant galaxies'' as centrals. The numbers and fractions 
of central/satellite galaxies in the three samples are listed in 
\autoref{tab:group_stat}. We find that the majority of the RR galaxies 
are centrals in their host groups, with a very high central 
fraction of 89\%, compared to 64\% and 76\% in the RN and SC samples, respectively. 
According to the group catalog, we find the median halo mass of 
the RR galaxies is $M_h\sim10^{12}h^{-1}\msun$, with $91\%$ of the 
sample (compared to $71\%$ and $84\%$ for the RN and SC samples) 
to be hosted by relatively low-mass halos with $M_h < 10^{13}h^{-1}\msun$. 
For the RN galaxies, their host halos span a wide range 
of halo mass, with a median halo mass of $M_h\sim10^{12.5}h^{-1}\msun$
and a significant fraction ($18\%$) above $M_h\sim10^{14}h^{-1}\msun$. 
This result clearly shows that, different from the RN galaxies, 
the RR galaxies are mainly central (or isolated) galaxies 
in relatively low mass halos, thus free of strong environmental effects
and more capable of accreting and/or retaining their cold gas. 

\begin{deluxetable}{cccc}[ht!]
	\tablenum{2}
	\tablecaption{Central/satellite numbers and fractions\label{tab:group_stat}}
	\tablewidth{0pt}
	\tablehead{
		\colhead{sample} & \colhead{centrals or isolated} & \colhead{satellites} & \colhead{Total}
	}
	\startdata
    RR & 42(89\%) & 5(11\%) & 47 \\ 
    RN & 98(64\%) & 56(36\%) & 154  \\ 
    SC & 134(76\%) & 43(24\%) & 177 
    \enddata
\end{deluxetable}

\begin{figure*}
    \centering
    \includegraphics[width=0.45\textwidth]{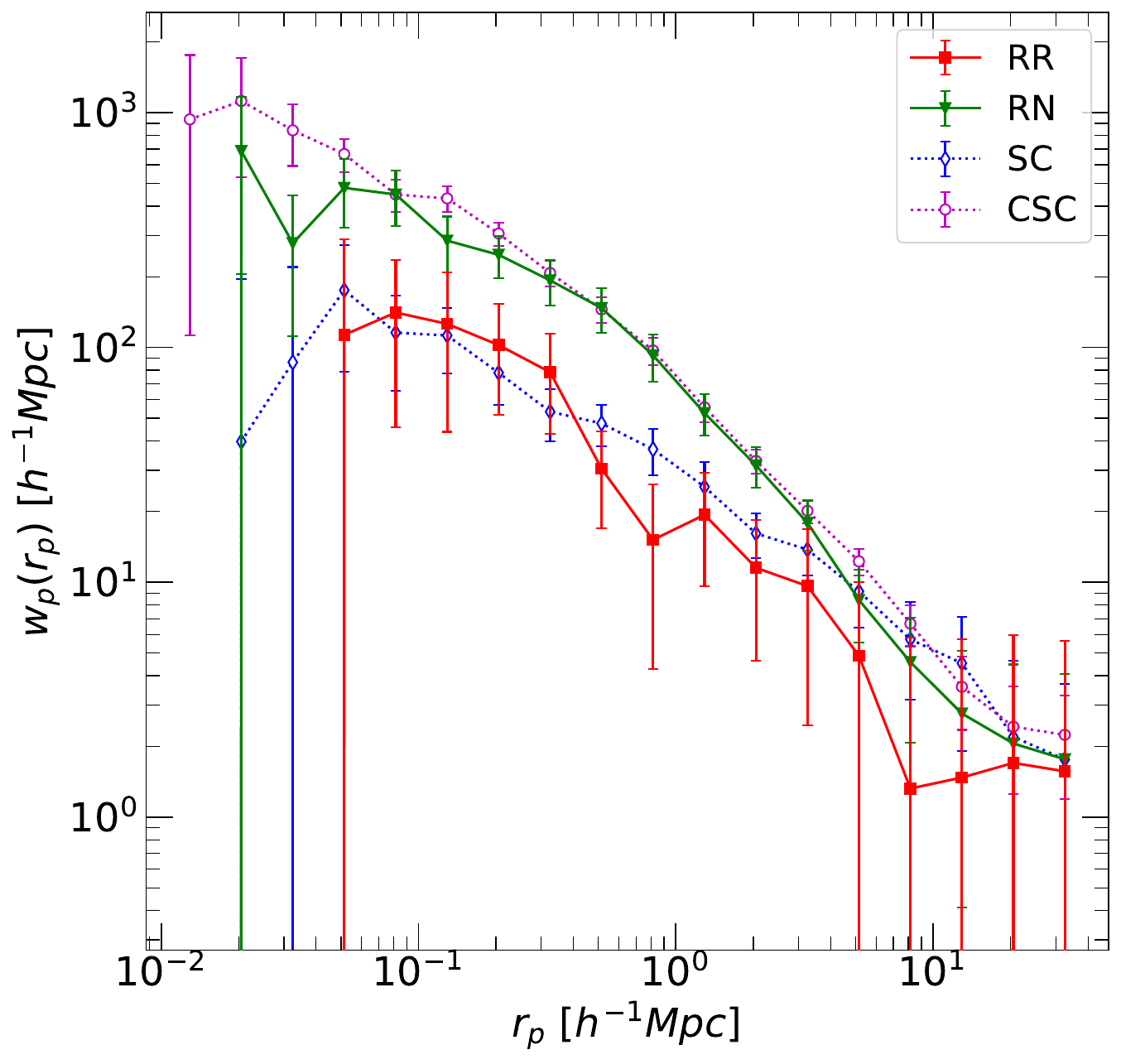}
    \includegraphics[width=0.45\textwidth]{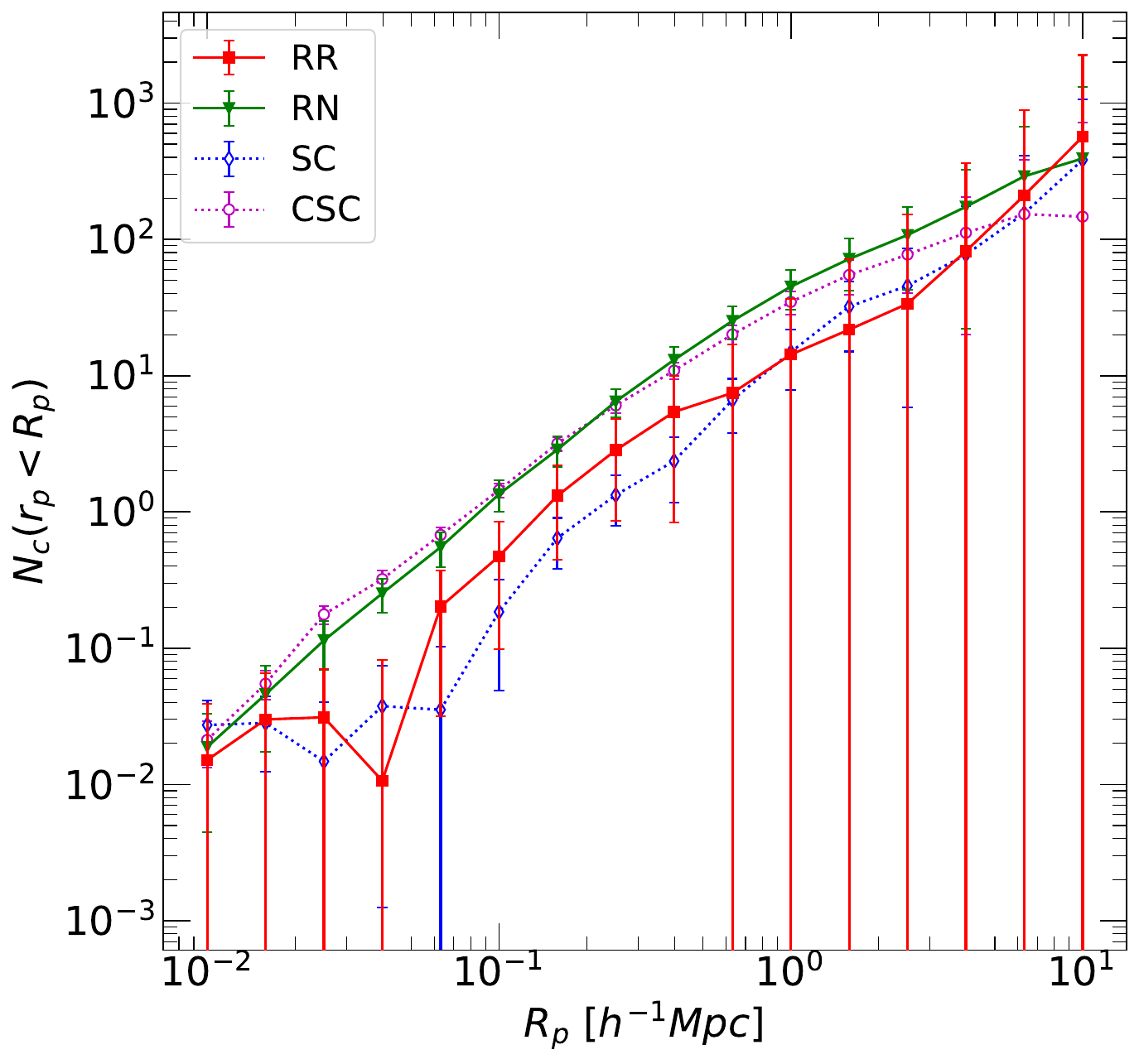}
    \caption{Measurements of projected cross-correlation 
    functions with respect to the SDSS reference sample
    (upper) and background-subtracted neighbor counts 
    in the SDSS photometric sample down to the $r$-band 
    magnitude limit of $r_{\rm lim}=21$ (lower).
    Results for the different samples are plotted 
    in different symbols/lines/colors, as indicated.}
    \label{fig:clustering}
\end{figure*}

Next, we examine the three-dimensional overdensity of the 
local environment of our galaxies, $\delta\equiv \rho/\bar{\rho}-1$, 
where $\rho$ and $\bar{\rho}$ are the local and mean 
matter density, respectively. Taken from the ``Exploring 
the Local Universe with the reConstructed Initial Density Field''
project \citep[ELUCID;][]{Wang2016}, the matter density field 
can well reproduce the distribution of both galaxies and 
groups of galaxies in the local Universe as observed by SDSS. 
Overdensities can be estimated at different scales by 
smoothing the density field with Gaussian kernels of different 
sizes. In \autoref{fig:overdensity} we compare the distribution
of the local overdensity as estimated at scales of 2 Mpc 
(left panel) and 4 Mpc (right panel) for different samples. 
At both scales, we find a trend for the RR galaxies to be located 
in relatively underdense regions compared to the RN sample, although the $p$-value of the two-sample Kolmogorov-Smirnov test shows that this difference is not very significant.
The difference is more pronounced at 4 Mpc than at 2 Mpc, as indicated by the $p$-values. 

We estimate two more statistics to further probe the environment 
of our galaxies over a wider range of spatial scales.
The first statistic is the projected cross-correlation function 
(PCCF), $w_p(r_p)$, measured for each of our samples with 
respect to a reference sample of $\sim5.3\times10^5$ galaxies 
from the SDSS spectroscopic survey. This statistic quantifies 
the clustering of galaxies over scales from a few 
$\times10$ kpc up to a few $\times10$ Mpc. The other statistic
is the background-subtracted neighbor count, $N_c(<R_p)$, 
that is, the average number of neighboring galaxies within 
a projected distance $R_p$ around the galaxies in our samples, 
as estimated in the SDSS photometric sample down to the 
$r$-band limiting magnitude of $r_{\rm lim}=21$ mag. 
The contribution by uncorrelated background galaxies is 
statistically estimated and subtracted. Detailed descriptions 
of the two statistics and tests/applications can be found 
in \citet{Li2006-AGN,Li2008a,Li2008b} and \citet{Wang2019}. 
\autoref{fig:clustering} shows our measurements of  
$w_p$ and $N_c$ for the different samples.
First of all, for the CSC sample both the $w_p(r_p)$ and 
	$N_c$ measurements are in good agreement with those of 
	the RN sample, demonstrating again that the RN sample 
	is a random subset of the red population of their mass.

On scales larger than a few Mpc, the RN and SC samples 
present similar $w_p(r_p)$ amplitudes, implying similar 
dark matter masses of their host halos, while the RR sample 
shows weaker clustering (although with large errors). 
This result is consistent with both the smaller halo masses of the 
RR galaxies as found above from the SDSS group catalog
and the lower local densities found at 4 Mpc from the 
reconstructed density field. 
On intermediate scales from a few Mpc down to $\sim100$ kpc, 
the RR and SC samples show similar clustering amplitudes, 
which are significantly lower than that of the RN sample. 
The stronger clustering of the RN sample than the SC 
sample can be attributed to the different \nuvr\
colors of the two samples. Previous studies of galaxy 
clustering have shown that, at scales below 
a few Mpc, red galaxies are more strongly clustered than 
blue galaxies of similar masses \citep[e.g.][]{2006MNRAS.368...21L}.
Since the RR and RN samples are matched in both stellar 
mass and color, the weaker clustering of the RR sample 
at intermediate scales can be understood from the 
aforementioned higher fraction of central galaxies. 
As shown in \citet{Li2006-AGN} and \citet{Wang2019}, the 
dip in $w_p(r_p)$ at intermediate scales can well be 
reproduced by models in which a higher-than-average 
fraction of the galaxies are located at the centers 
of their host dark matter halos. 

\begin{figure*}
    \centering
    \includegraphics[width=0.9\textwidth]{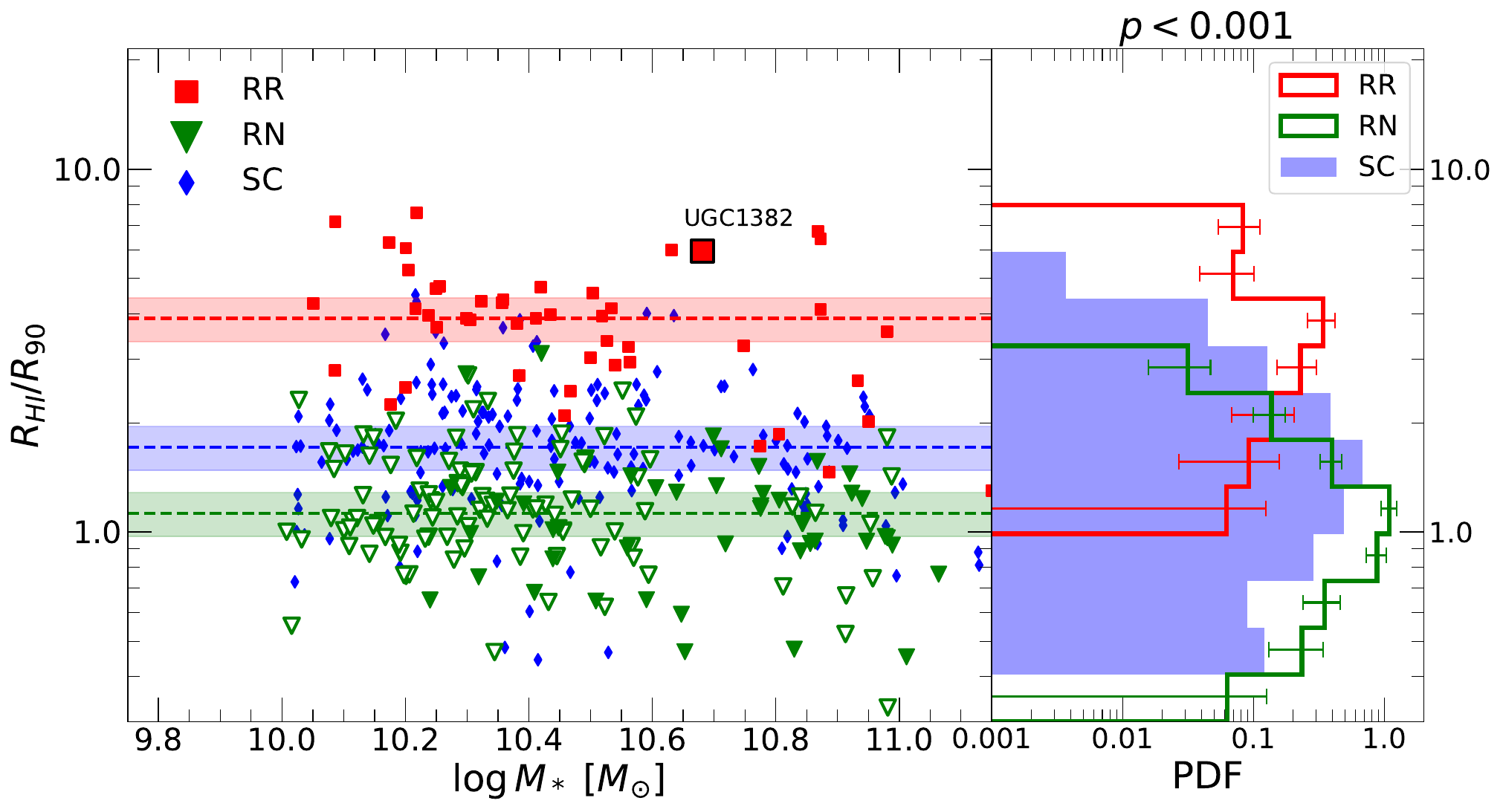}
    \caption{The \hi-to-optical radius ratio as a fucntion of stellar mass. The red squares represent the RR sample.
    The green triangles represent the RN sample. The blue diamonds represent the SC sample. Solid and open symbols represent \hi\ detections and non-detections (upper limits) separately. The dashed lines show the median value
    of the \hi-to-optical radius ratio for the three samples, with the shaded region indicating the uncertainty due to the scatter of
    the \hi\ size-mass relation. The histograms show the normalized distribution of the \hi-to-optical radius ratio. 
    The p-value of the two-sample KS test conducted between the RR and the RN sample is indicated on top of the left panel.
   The red square is UGC 1382 (id=0 in Table~\ref{tbl:galaxylist}), the only galaxy in the RR sample
   that has resolved \hi\ observations available for measuring \hi\ disk size \citep{Hagen2016}.}
    \label{fig:HI_to_opt_size}
\end{figure*}

The $N_c$ measurements present similar behaviors to 
$w_p(r_p)$. This is expected 
as the two statistics are closely related. Compared to 
$w_p(r_p)$, the $N_c$ are measured with smaller 
errors on the smallest scales ($\lesssim100$kpc) thanks 
to the much deeper photometric sample. We see a trend 
for a dip at $R_p\sim40 h^{-1}$kpc as well as an upturn 
at smaller scales in the average number of close companions 
around the RR galaxies. This result might indicate some 
role of tidal interactions or mergers for RR galaxies. 
Large samples and more studies are needed to test 
this hypothesis. 

\subsection[short]{\hi-to-optical size ratio}

Previous studies have revealed a tight relation between the 
\hi\ disk size and \hi\ mass for nearby galaxies 
\citep[e.g.][and reference therein]{2016MNRAS.460.2143W}. 
The unusually high \hi\ mass fractions of the RR galaxies 
thus imply large \hi\ disk sizes relative to galaxies with 
normal \hi\ contents, if we assume that different types of
galaxies follow the same \hi\ size-mass relation. 
We estimate the radius of the \hi\ disk ($R_{\rm HI}$) for 
each galaxy in our samples using the \hi\ disk size-mass 
relation from \cite{2016MNRAS.460.2143W}, where $R_{\rm HI}$ 
is defined as the radius at which the 
surface \hi\ mass density is equal to $\rm 1\ \msun\ pc^{-2}$. 
The left-hand panel of \autoref{fig:HI_to_opt_size} shows the 
ratio of the estimated $R_{\rm HI}$ to the optical radius 
$R_{90}$ as a function of stellar mass for the different 
samples of galaxies, with the side panel showing the histogram 
of $R_{\rm HI}/R_{90}$ of all the galaxies in each sample. 
As expected, the RR galaxies have the largest \hi-to-optical 
size ratios with an average of $R_{\rm HI}/R_{90}\sim$4, 
compared to $R_{\rm HI}/R_{90}\sim$1 for the RN sample 
and $R_{\rm HI}/R_{90}\sim$1.7 for the SC sample. We see 
weak dependence on the stellar mass for all the three samples, 
although there is considerable scatter at fixed mass.
Note that the RN sample is dominated by \hi\ upper limits,
and so the estimated \hi\ sizes are also upper limits. This 
means that the \hi\ disk of red galaxies with normal \hi\ 
content is comparable or even smaller than their optical disk. 

One of our RR galaxies,
UGC 1382 (id=0 in Table~\ref{tbl:galaxylist}) with resolved \hi\ 
observations available for measuring \hi\ disk size \citep{Hagen2016} 
is highlighted in \autoref{fig:HI_to_opt_size} as a red square with black borders. 
This galaxy is above the average \hi\ size-mass relation from 
\citet{2016MNRAS.460.2143W}, but with a deviation that is only slightly larger 
than the 3$\sigma$ scatter. Although with only one galaxy, 
this result supports the assumption that the galaxies from 
different samples follow the same \hi\ size-mass relation. 
This assumption is actually consistent with previous studies 
of resolved \hi\ observations of nearby galaxies. For 
instance, using \hi\ maps of nearby galaxies obtained from 
the Westerbork Synthesis Radio Telescope (WSRT), 
\cite{2013MNRAS.433..270W} found that their ``bluedisk'' galaxies with 
unusually high \hi\ mass fractions lie on the same 
\hi\ mass versus \hi\ size relation as the control 
galaxies with normal \hi\ content, although the former 
have much larger \hi-to-optical size ratios, 
with $R_{\rm HI}$ extending to values as large as $\sim100$kpc
for \mhi$\sim2\times10^{10}$\msun\ 
(see Figure 6 of their paper). Therefore, applying the 
same size-mass relation to estimate \hi\ radii for different 
samples appears to be a reasonable choice for our analysis
here, before large samples of resolved \hi\ observation 
become available in the future.

\begin{figure*}
    \centering
    \includegraphics[width=0.495\textwidth]{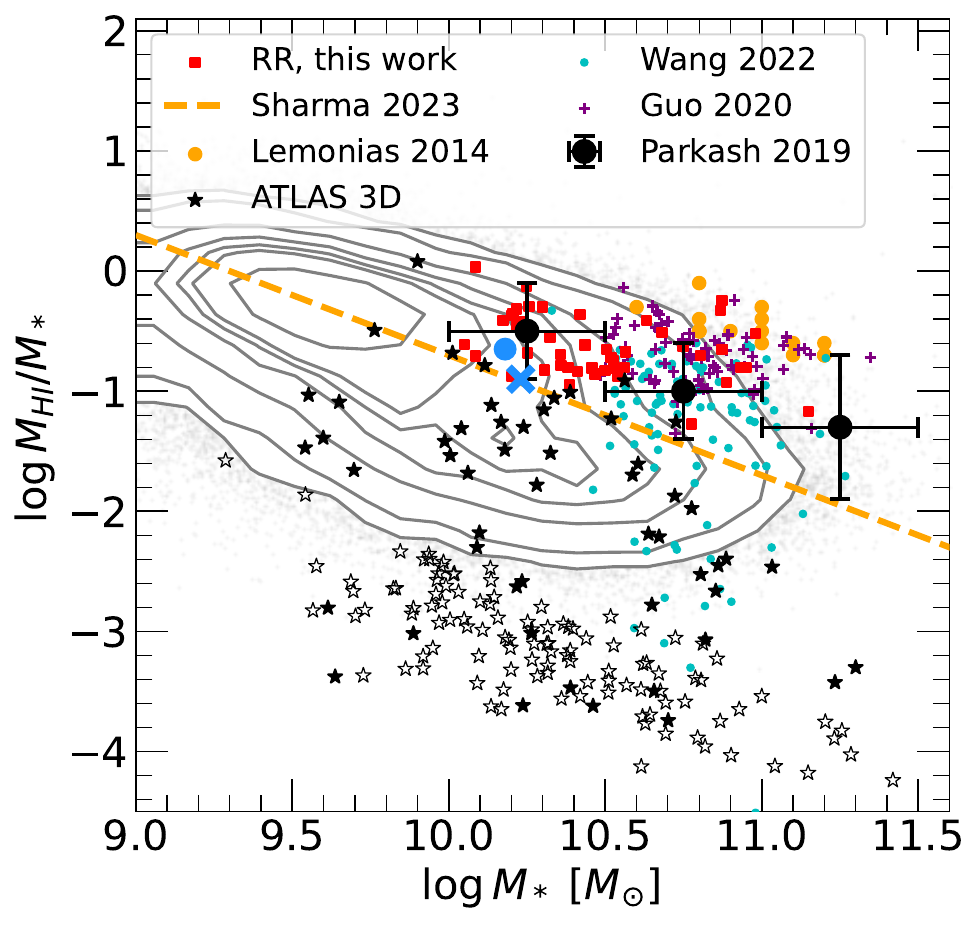}
    \includegraphics[width=0.495\textwidth]{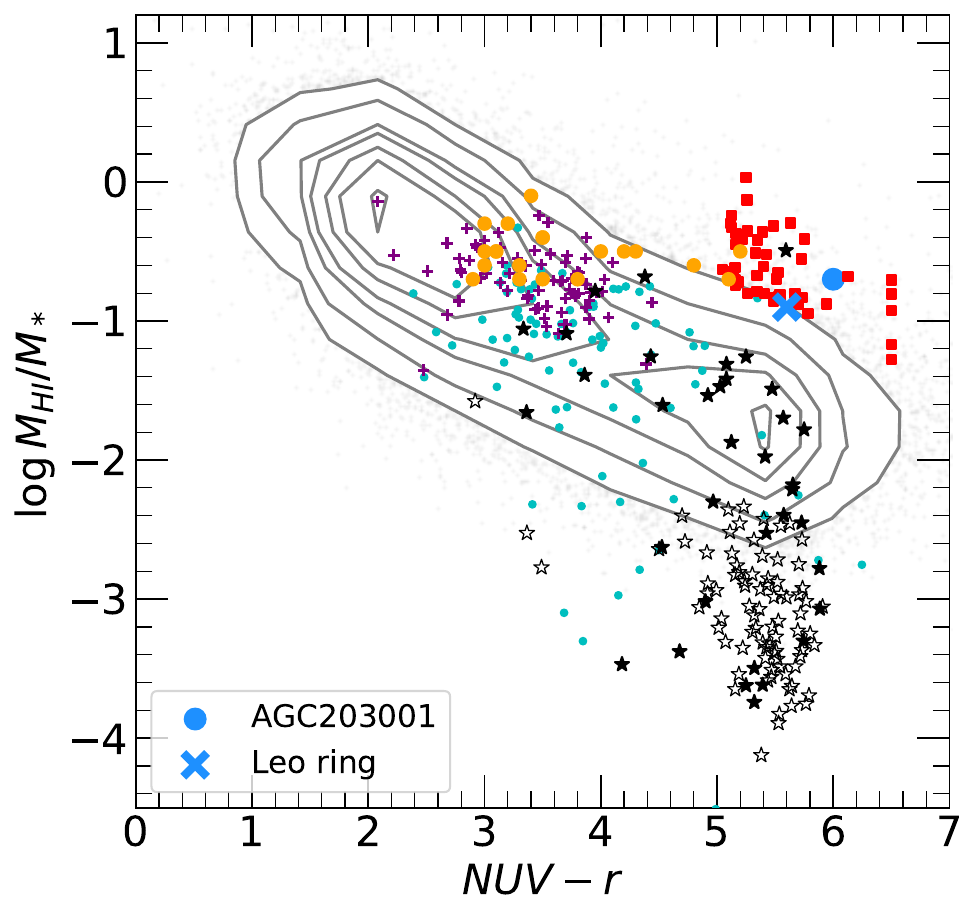}
    \includegraphics[width=0.495\textwidth]{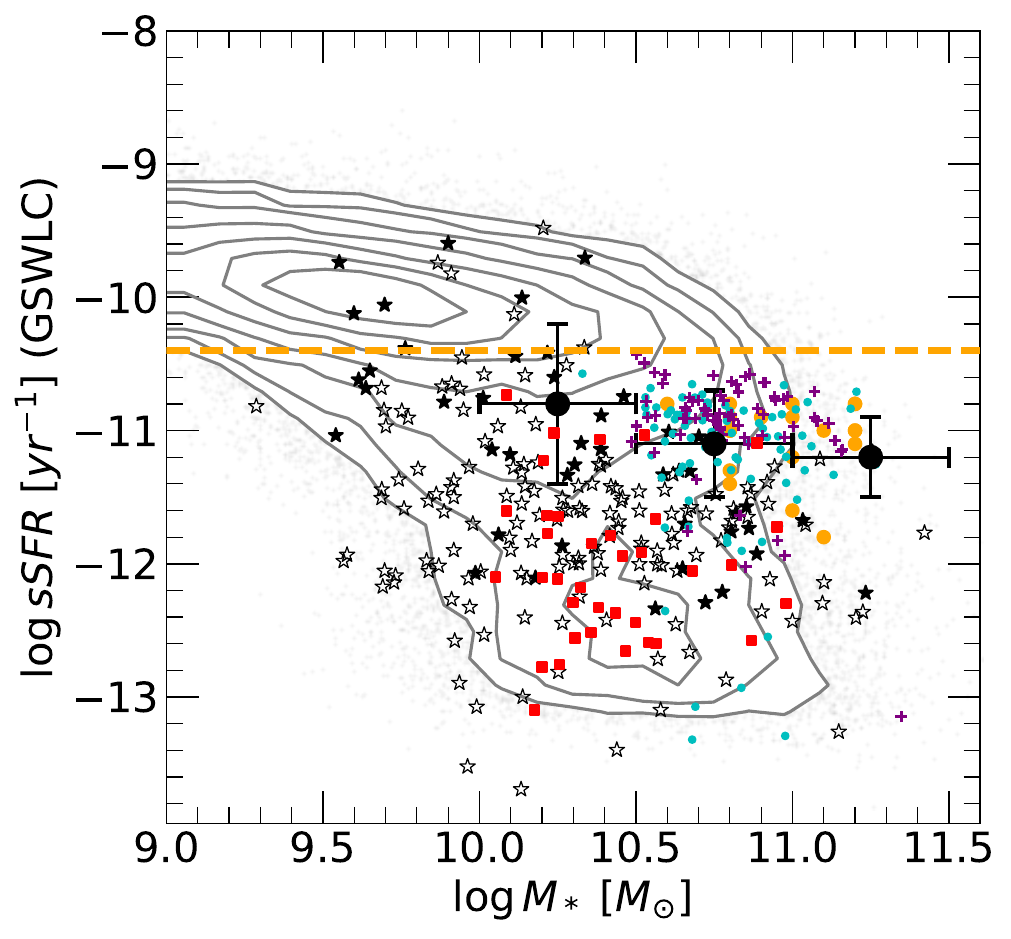}
    \includegraphics[width=0.495\textwidth]{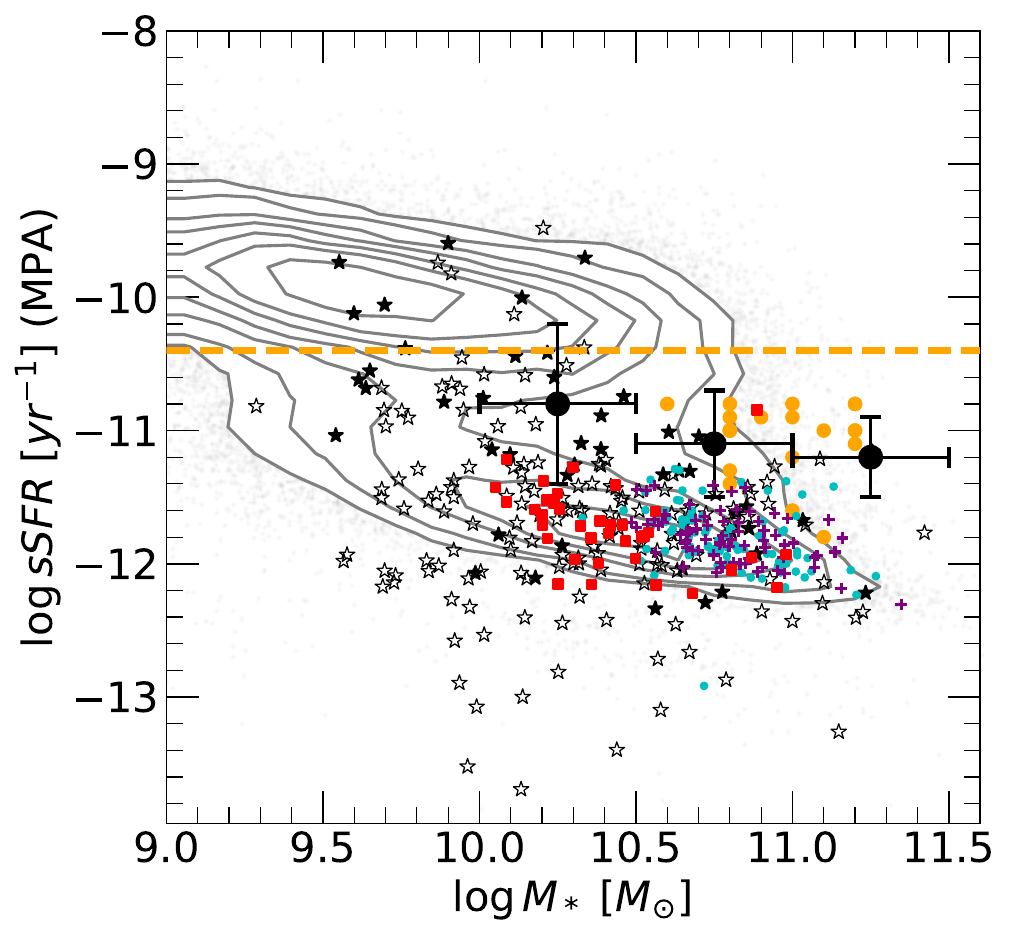}    
    \caption{Comparisons of the RR sample studied in this work 
    with previous \hi\ samples in the diagrams 
    of \mhi/\mstar\ versus \mstar\ (top-left), 
    \mhi/\mstar\ versus \nuvr\ (top-right), and 
    sSFR versus \mstar\ (with SFRs from GSWLC in the bottom-left 
    panel, and SFRs from the MPA/JHU catalog in the bottom-right panel). 
    Solid and open symbols represent \hi\ detections and non-detections (upper limits) separately.
    Red squares are the RR galaxies. 
    Yellow dots are the \hi-rich 
    galaxies with suppressed star formation from \citet[][]{Lemonias2014}. 
    Purple pluses and cyan dots are massive red spirals from 
    \citet{Guo2020} and \citet{Wang2022-FAST}. 
    Black solid dots with error bars are the median relation and scatter of the 
    sample from \citet{Parkash2019}.
    The yellow dashed lines indicate the sample selection 
    in \citet{Sharma2023}: \mhi$>10^{9.3}$\msun\ (top-left)
    and sSFR$<10^{-10.4}$yr$^{-1}$ (bottom panels). 
    Black solid and open stars represent the \hi\ detections and 
	non-detections (upper limits) of the ATLAS$^{\rm 3D}$ sample, for which 
	stellar masses are derived from $\mathrm{M_{JAM}}$ given by \cite{Cappellari2013} based on 
	dynamical modeling and sSFRs are estimated from PAH luminosities 
	by \cite{Kokusho2017} based on SED fits to data from AKARI, WISE and 2MASS.
	The blue cross and dot indicate the Leo ring and AGC203001
    	for which the \hi\ and stellar masses and the colors are taken from 
    	\citet{Bait2020}. The grey contours represent the SDSS volume-limited sample, 
    	of which the \hi\ mass of each galaxy is predicted by the estimator from \cite{XiaoLi}.}
    \label{fig:comparisons}
\end{figure*}

\section[short]{Discussion} \label{sec:discussion}

\subsection{Uniqueness of the RR galaxies}

In this subsection we compare our RR sample with similar 
samples in previous studies. First, we consider the 
sample of early-type galaxies (ETGs) from the ATLAS$^{\rm 3D}$ 
\hi\ survey by \citet{Serra2012-ATLAS3D}. 
Similar to the RR galaxies studied here, the ETGs in ATLAS$^{\rm 3D}$ 
	also presented much larger \hi\ discs than their optical sizes and 
	the \hi-richest ETGs were also found in low density environments
	\citep{Serra2012-ATLAS3D}.  
	For quantitative comparison, we have 
	attempted to match the ATLAS$^{\rm 3D}$ 
	sample with our SDSS galaxy catalogs. Due to the relatively low redshifts 
	of the ATLAS$^{\rm 3D}$ sample, none of the galaxies are included in 
	GSWLC and only a fraction have counterparts in the NSA and the MPA/JHU catalog.
Therefore, for the ATLAS$^{\rm 3D}$ galaxies, we opt for stellar masses derived by \cite{Cappellari2013} based on dynamical modeling, i.e. $M_{\rm JAM}$ in that paper. By comparing $M_{\rm JAM}$ with the NSA $M_{\ast}$ for those galaxies that have counterparts in NSA, we find $M_{\rm JAM}$ is systematically larger by $\sim0.3$ dex than $M_{\ast}$ with no obvious dependence on $M_{\rm JAM}$. To take into account this difference, we use $\log_{10}M_{\rm JAM}-0.3$ as the stellar mass for the ATLAS$^{\rm 3D}$.
 The SFRs are estimated by \cite{Kokusho2017} from PAH luminosities 
	which were obtained from SED fitting to data from AKARI, WISE and 2MASS. 
	We have compared the SFRs for those with counterparts 
	in the MPA/JHU catalog and found no systematic differences. 
	In \autoref{fig:comparisons} we compare the ATLAS$^{\rm 3D}$ sample with 
	the RR sample in terms of the distributions of \mstar, \mhi/\mstar, \nuvr\ and sSFR. 
We find that, when compared to the RR sample, the ETGs 
from ATLAS$^{\rm 3D}$ mostly have similarly red colors and low sSFRs 
as expected, and significantly lower \hi\ mass fractions. 
Only one galaxy from the ATLAS$^{\rm 3D}$ meets our selection criteria of the RR sample,
while most of them fall on or even below the contours of the general 
red population, indicating that the majority of red early-type galaxies 
are \hi\ normal or \hi\ poor, as generally expected.
The difference between the RR sample and the general ETG 
population can also be seen from the bottom-right panel 
in \autoref{fig:diagrams}, where the RR, RN and CSC samples 
show very similar distributions in T-type, all dominated 
by early-types, but the RR sample is very different in \hi\ mass 
fraction to the other two samples.
The E+A galaxies previously studied are also different from 
our RR galaxies. In fact, Figure 3 of \citet{Zwaan2013-PSB} 
has shown that their E+A galaxies as well as the several from 
earlier studies are mostly ``green'' with $3<~$\nuvr$~<4$ and 
\hi-normal for their color, with $\log_{10}$(\mhi/\mstar)$~<-1$. 

The samples of \hi-rich galaxies with suppressed or low star 
formation \citep{Lemonias2014,Parkash2019,Sharma2023} 
and massive red spirals \citep{Guo2020,Wang2022-FAST} are 
more similar to our RR sample. The selection criteria of these 
previous samples are mentioned in \autoref{sec:intro}.
The comparison of these samples with the RR sample is 
shown in \autoref{fig:comparisons}.
As can be seen from the top-left panel, all the samples in 
comparison are limited to relatively high stellar mass and 
high \hi\ mass fraction with respect to the complete SDSS sample, 
although the FAST sample from \citet{Wang2022-FAST} includes 
some galaxies with low \hi\ mass fractions.
Our RR sample is distinct from other samples by having the 
reddest colors (top-right panel) and the lowest sSFRs 
(bottom-left panel), both indicating the fully 
quenched status of star formation in these galaxies. 
In contrast, both the \hi-rich galaxies with suppressed 
star formation from \citet{Lemonias2014} and the massive 
red spirals from \citet{Guo2020} and \citet{Wang2022-FAST} 
are mostly ``green'' with intermediate colors, 
$3\lesssim~$\nuvr$~\lesssim5$, and consistently they fall 
mainly in the transition region between the star-forming 
and quiescent sequences (bottom-left panel), indicating 
that these galaxies are not fully quenched. As pointed out 
in \citet{Zhou2021}, red spirals selected by optical colors 
such as $u-r$ are mostly green/blue in \nuvr, as the UV 
emission is more sensitive than optical colors to the ongoing 
weak star formation in galaxies. The fact that the sample 
of \citet{Parkash2019} is also limited to the transition 
region suggests that the criterion of mid-infrared color 
(${\rm W2}-{\rm W3}<2$) is unable to select fully quenched 
galaxies as well. \citet{Sharma2023} did not tabulate their 
galaxies, but it is natural to expect their sample to occupy 
similar regions in the sSFR versus \mstar\ diagram considering 
the same color selection as adopted in \citet{Parkash2019}.  
We note that massive red spirals show quite different 
distributions in the sSFR versus \mstar\ diagram depending 
on the adopted SFR estimates. When the MPA/JHU catalog is used  
instead of GSWLC, these galaxies move from the transition 
region to the quiescent sequence, showing similarly low sSFRs to 
our RR galaxies. We argue that this discrepancy 
largely attributes to the larger uncertainties and biases 
in the SFRs from the MPA/JHU catalog than from the GSWLC at 
intermediate-to-low SFRs (see \citealt{Li-Ho-2023} for a recent comprehensive investigation 
and discussion). 

\begin{figure*}
    \centering
    \includegraphics[width=0.45\textwidth]{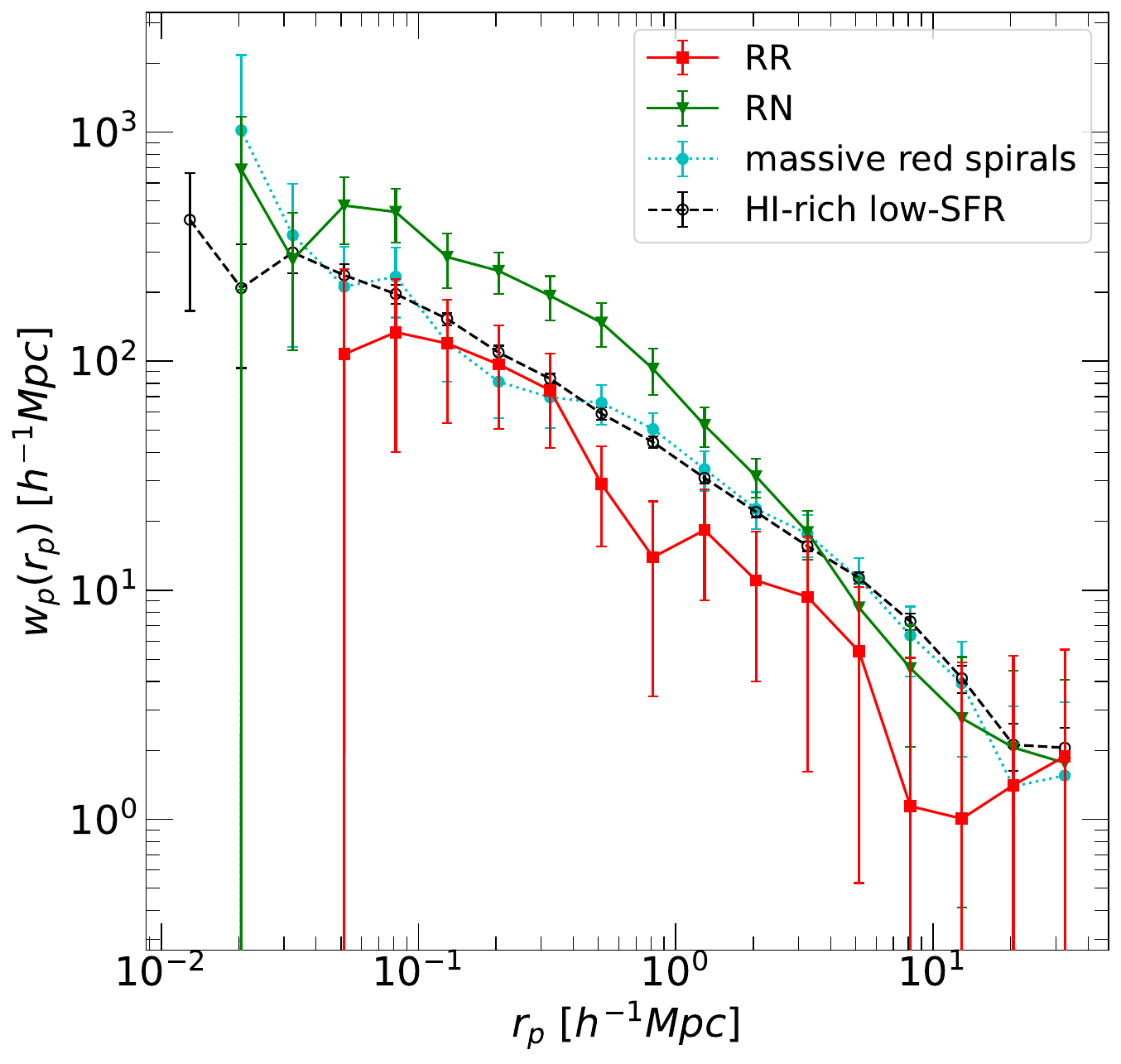}
    \includegraphics[width=0.45\textwidth]{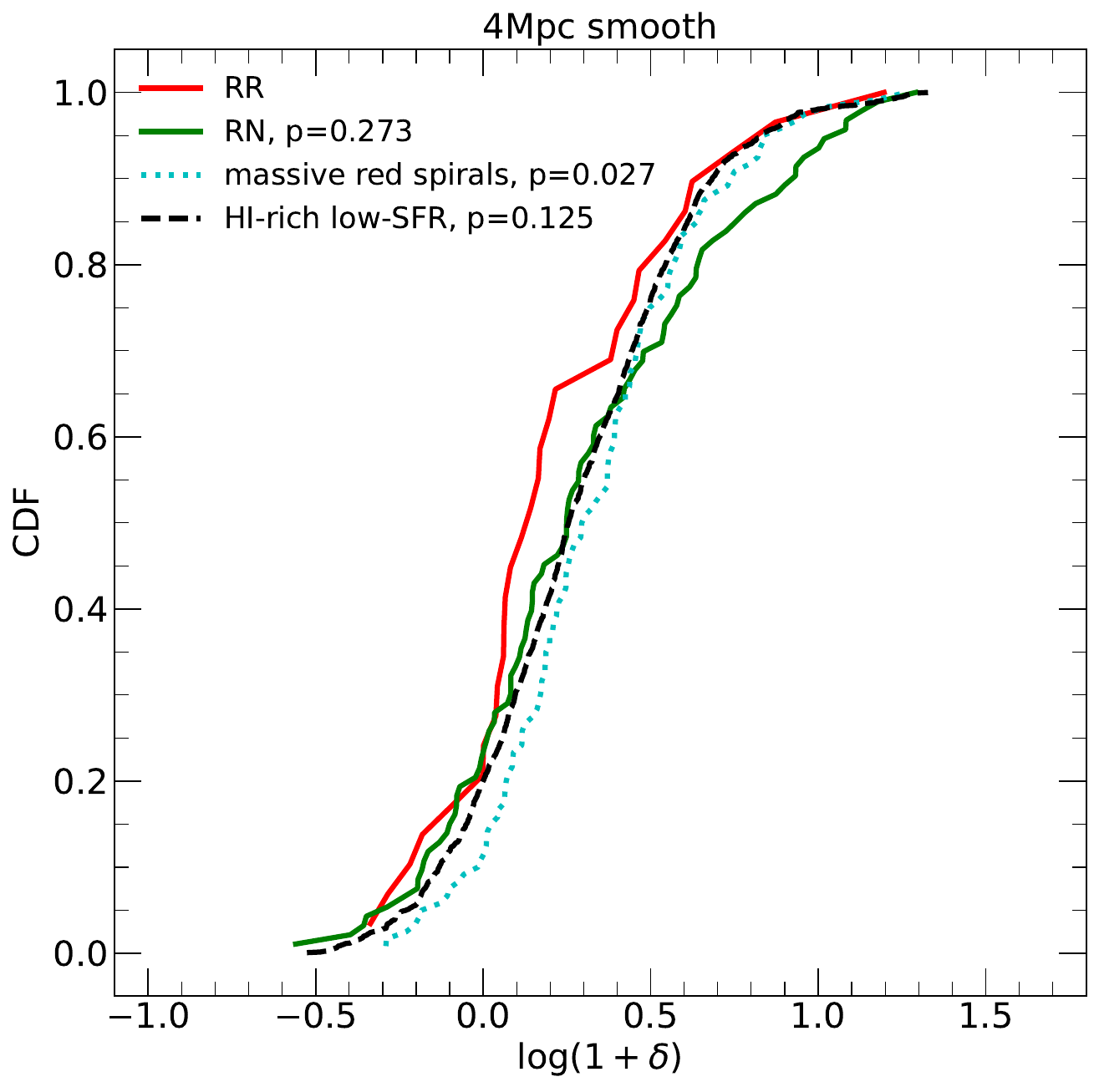}
    \caption{Projected two-point correlation function $w_p(r_p)$ 
    (left) and cumulative distributions of local density estimated 
    at 4 Mpc (right), as measured for 
    the RR and RN samples studied in this work, as well as 
    the \hi-detected massive red spirals from \citet{Guo2020} 
    and \citet{Wang2022-FAST} and a sample selected from 
    SDSS to mimic the \hi-rich low-SFR samples studied in 
    previous studies (see the text for details). The value of "p" in the legend of the right panel is the p-value of the two-sample KS test conducted between the RR and the corresponding sample.}
    \label{fig:compare_environment}
\end{figure*}

We conclude that the RR sample selected in the \mhi/\mstar\ 
versus \nuvr\ diagram represents a unique population of 
galaxies, which are fully quenched but contain unusually 
large amounts of \hi\ gas. Previous samples selected 
by $u-r$ \citep{Guo2020,Wang2022-FAST}, or ${\rm W2}-{\rm W3}$
\citep{Parkash2019,Sharma2023}, or by a fixed cut in sSFR 
\citep{Lemonias2014} represent galaxies in the ``green valley'',
i.e. in the transition phase to being fully quenched. 
Their \hi\ contents can be partly (if not fully) explained 
by the blue/green \nuvr\ colors and the intermediate sSFRs. 

Considering that the cessation of star formation in massive 
galaxies happens from inside out \citep[e.g.][]{Li2015-insideout,Wang2018-insideout},
it could be that the \hi-rich galaxies in previous 
samples and the RR galaxies in our sample have the same 
origin and follow a common evolutionary path, but they 
are at different stages. In this case, the RR sample and 
the previous samples should share some common properties. 
In fact, like our RR galaxies, the \hi-rich galaxies 
in \citet{Lutz2018,Lutz2020} were also found to have similar 
optical properties (e.g. the size-mass relation and radial 
distributions of gas metallicity) to that of 
\hi-normal galaxies. In addition, the large \hi-to-optical 
size ratios found for our RR galaxies was also found in 
\citet{Lemonias2014} with the help of resolved \hi\ 
observations: most of the massive \hi-rich galaxies with 
suppressed star formation in their sample had \hi\ radii 
at least twice as large as their optical radii (see their 
Figure 7). The authors found all of their galaxies had 
substantially low \hi\ surface densities, likely to be the 
main cause of their low sSFRs. Given the even larger 
\hi-to-optical radius ratios of our RR galaxies, they
are expected to have even lower \hi\ surface densities. 

On the other hand, however, our RR galaxies also show quite 
different properties in many other aspects from galaxies used in previous
investigations. For instance, 
\citet{Lemonias2014} found some evidence that AGN or bulges 
also contributed to the star formation suppression in their 
galaxies, giving support to AGN feedback and morphology quenching 
\citep{Martig2009}. \citet{Sharma2023} 
also found some evidence for AGN feedback and bar quenching 
for the \hi-rich but low-SFR galaxies based on the IFS 
data from MaNGA. 
In addition, \citet{Parkash2019} found a high fraction 
(75\%) of their galaxies to be LINERs or LIERs, and they 
speculated that the presence of \hi\ is a precondition for 
LI(N)ER emission in galaxies. \citet{Sharma2023} 
found a similarly high fraction ($77\pm11\%$) of LIERs. 
However, they found their control sample of low-\hi\ and 
low-SFR galaxies also has a high fraction of LIERs 
($60\pm10\%$), and concluded that \hi\ content is not 
required for the high LIER fraction to correlate with the low SFR.
We find no evidence for either morphology quenching, given 
the similar distributions in $B/T$ and $\sigma_\ast$ for 
the RR, RN and CSC samples as seen in \autoref{fig:diagrams}, 
or bar quenching given the similar distributions in 
$P_{\rm bar}$ as seen in \autoref{fig:histograms}.
Selected from SDSS, most of our RR galaxies do not have 
integral field spectroscopy, but we have 
examined their single-fiber SDSS spectra. We find that the 
emission lines relevant for AGN identification are weak 
in most galaxies, indicative of no AGN or LIER in their 
central regions. This result is consistent with the global 
quiescence of the RR galaxies. 

In \autoref{fig:compare_environment} we compare the 
projected 2PCF and local densities between our 
RR sample and the previous samples. The samples from 
\citet{Lemonias2014} and \cite{Parkash2019} are 
too small to obtain reliable measurements, while 
\citet{Sharma2023} does not tabulate their galaxies. 
In order to compare, we select a sample of 
``\hi-rich low-SFR'' galaxies from the SDSS
to mimic their samples, by applying the following two 
criteria: $\log_{10}$\mhi$~>9.3$ \&\& 
$-11.5<\log_{10}({\rm sSFR})<-10.4$. We further require 
this sample to have the same stellar mass distribution 
as our RR sample. The sample includes 2,200 galaxies.
The measurements for this sample are plotted as black 
symbols in the figure, and those of a combined sample 
of massive red spirals with \hi\ detections from ALFALFA 
\citep{Guo2020} and FAST \citep{Wang2022-FAST} are 
plotted as cyan symbols. As can be seen, 
the ``\hi-rich low SFR'' sample and the massive red spirals
are very similar in both $w_p(r_p)$ and $\log_{10}(1+\delta)$, 
but when compared to them, the RR galaxies are less 
clustered at all scales except $r_p\sim 0.2-0.3$ $h^{-1}\rm Mpc$ 
and have lower densities on average. We have also 
examined the central and satellite fractions of the 
two samples based on the SDSS group catalog, and found 
a central fraction of 73\% for the massive red spiral 
galaxies and 70\% for the ``\hi-rich low SFR'' sample.
These results imply that low-density environments and 
high central fractions are necessary conditions to make
the RR galaxies a unique population, which is unlikely 
to entirely follow the same evolutionary pathway as 
the \hi-rich low-SFR samples previously studied.

\subsection{Existence and rareness of the RR galaxies} \label{subsec:HI_fraction}

In a recent review, based on the xGASS sample \citet{Saintoge2022}
concluded that there is no evidence for a significant population 
of passive galaxies with \hi\ reservoirs comparable to those 
of star-forming main-sequence galaxies. Our result is not in 
contradiction with theirs. As can been seen from 
\autoref{fig:sample_selection}, xGASS indeed contributes only 
a few galaxies to our RR sample. In fact, our RR sample is contributed 
mainly by ALFALFA, which is much larger in sample size though 
less complete in \hi\ mass fraction when compared to xGASS. 
This strongly suggests that the quenched but \hi-rich galaxies 
are rather rare, and they can be identified as a significant 
population for statistical studies as done in this work only if 
the parent sample is large enough. The rareness of the RR galaxies 
can be more clearly seen when compared to the SDSS volume-limited 
sample. As we have seen from \autoref{fig:fHI_vs_SFR}, the 
majority of our RR galaxies are found beyond the 95\% contour 
of the SDSS sample. In addition, as can be seen in \autoref{fig:comparisons},
	all the previously studied samples including the \hi-rich but low SFR 
	sample \cite{Lemonias2014}, the massive red spirals  \citep{Guo2020, Wang2022-FAST}, 
	and the ATLAS$^{\rm 3D}$ sample \citep{Serra2012-ATLAS3D}
	contain a few galaxies satisfying our criteria of RR galaxies. 
	This result is also consistent with both the existence and the rareness 
	of the RR galaxies.

\begin{figure*}
    \centering
    \includegraphics[width=\textwidth]{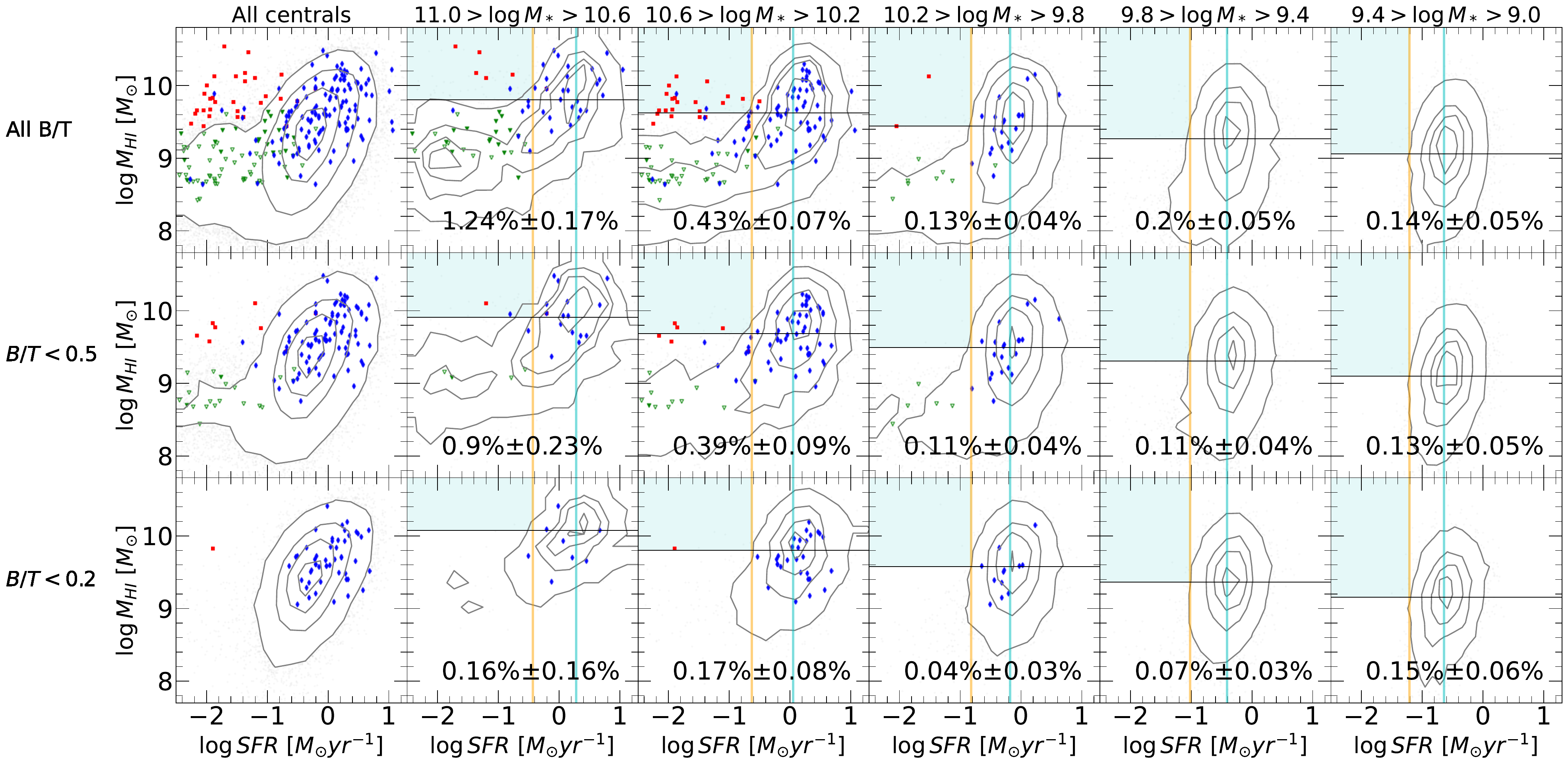}
    \caption{Distributions of central galaxies from the RR, RN, SC and 
    SDSS samples on the \mhi\ versus SFR diagram. Panels from left to 
    right correspond to different stellar mass ranges as indicated 
    above the top panels, while panels from top to bottom correspond to 
    different thresholds of the bulge-to-total ratio as indicated on the 
    left-hand side. In each panel, the red, green and blue symbols 
    represent the RR, RN and SC galaxies, while the black contours 
    enclose 20\%, 45\%, 70\% and 95\% of the SDSS sample. The cyan, 
    yellow and black lines indicate the SFR on the best-fitting 
    star-forming main sequence (SFMS), the $2\sigma$ lower limit of 
    the SFMS and the median \hi\ mass of the main-sequence galaxies 
    (those falling within $1\sigma$ around the SFMS). The fraction of 
    \hi-rich but low-SFR galaxies in the SDSS sample (i.e. those falling 
    in the cyan region) and its Poisson error are indicated in each panel.}
    \label{fig:mhi_sfr}
\end{figure*}

\begin{figure*}
    \centering
    \includegraphics[width=\textwidth]{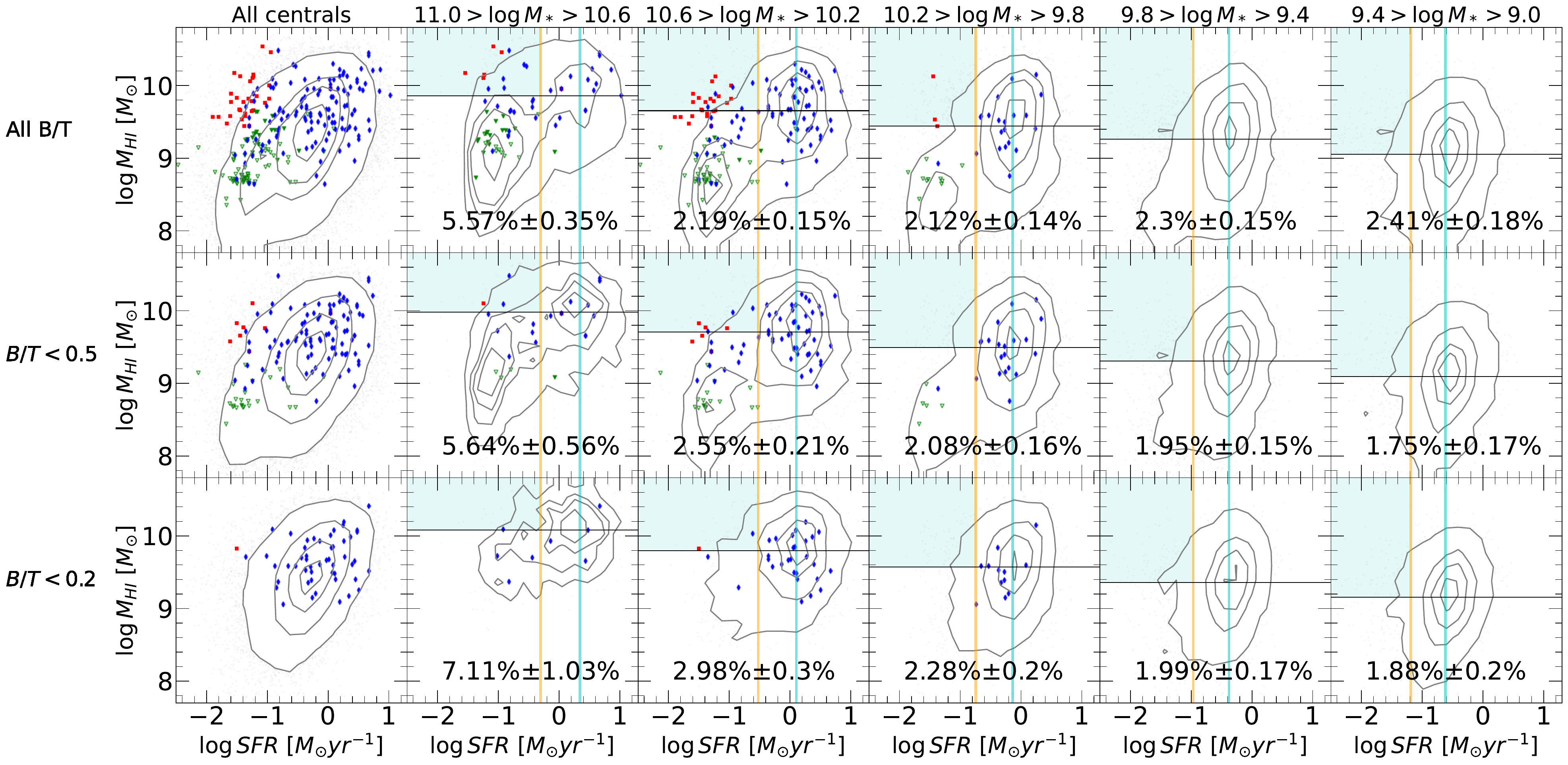}
    \caption{Same as \autoref{fig:mhi_sfr} except that the SFR estimates
    are take from the MPA/JHU catalog instead of GSWLC.}
    \label{fig:mhi_sfr_mpa}
\end{figure*}

Using data from both xGASS and ALFALFA, however, \citet{Chengpeng19} 
claimed that nearly all the massive quiescent central disc galaxies
are \hi-rich, with \mhi\ similar to that of star-forming galaxies. 
This result was questioned by \citet{Cortese2020} who found 
no passive disc galaxies in xGASS with \hi\ reservoirs comparable 
to those typical of star-forming galaxies. The authors suggested 
that the previous claim by \citet{Chengpeng19} was due to the use 
of aperture-corrected SFR estimates from the MPA/JHU catalog,
in which the global star formation of \hi-rich galaxies with 
extended star-forming discs is significantly underestimated. 
Here we revisit this problem taking advantage of the SDSS
volume-limited sample for which we have estimated an \hi\ mass 
for each galaxy. Adopting the same sample selection as in 
\citet{Chengpeng19}, we select a sample of 1609 massive 
($10^{10.6}M_{\odot}<M_*<10^{11}M_{\odot}$) central disk 
galaxies using the SDSS group catalog \citep{2007ApJ...671..153Y} 
and the SDSS morphology catalog from \citet{2018MNRAS.476.3661D}.
We define central galaxies as the most massive galaxy of their 
group and select disk galaxies by requiring T-type$~>0$ and 
$P_{\rm disk}>0.5$, where $P_{\rm disk}$ is the probability for a 
galaxy to be a disk provided by the morphology catalog. The average 
and scatter of \mhi/\mstar\ and \mhi\ of this sample as functions 
of \nuvr\ and SFR are plotted as the orange lines and shaded regions 
in \autoref{fig:fHI_vs_SFR}. We see the massive central disc 
galaxies to roughly follow the SC and RN samples at the high-
and low-SFR ends in terms of \mhi/\mstar, thus with much lower 
\hi\ mass fractions than the RR galaxies of similar SFRs. 
Because of their higher stellar masses, these galaxies have 
higher \mhi\ than the RN and SC samples at a given \nuvr\ or 
SFR, but still have lower \mhi\ than the RR galaxies. This result 
suggests both that the massive quiescent central disk galaxies 
investigated in \citet{Chengpeng19} are \hi-poor as expected 
from their red colors and low SFRs, and that these galaxies 
are different from the RR galaxies in our sample which are 
truly quenched, have unusually high \hi\ mass fractions, but 
are rare.

Our result is consistent with \cite{Cortese2020} where 
both \hi\ detections and non-detections from xGASS were used 
to quantify the \hi\ reservoirs of central galaxies as a function 
of stellar mass, SFR and $B/T$. Following the format of their 
Figs. 1 and 2 but using our volume-limited SDSS sample instead 
of xGASS, we examine the \mhi--SFR relationship for central galaxies 
in different intervals of \mstar\ and below different $B/T$ 
upper limits, plotted as black contours in 
\autoref{fig:mhi_sfr} (with SFRs from GSWLC) and 
\autoref{fig:mhi_sfr_mpa} (with SFRs from the MPA/JHU catalog). 
In each panel, we indicate the best-fitting star-forming main 
sequence (SFMS) and the 2$\sigma$ lower limit of the SFMS, 
as well as the median \hi\ mass of the SFMS galaxies (those 
falling within $1\sigma$ around the SFMS). The low-SFR but 
\hi-rich galaxies are then 
defined to be those with SFR below the 2$\sigma$ lower limit 
of the SFMS and above the median \mhi, as indicated by the cyan 
area. We calculate the fraction of the SDSS galaxies falling 
in this area and indicate the result in each panel. As can be 
seen, the fractions are small
in both figures and in all panels. In the case with the GSWLC 
SFRs, the fraction is at most $(1.24\pm0.17)\%$ 
as found for all the centrals in the most massive bin, and 
decreases as one goes to galaxies with lower \mstar\ and 
smaller $B/T$. Although the fractions become larger when  
the MPA/JHU SFRs are used, this type of galaxy is still a 
minor population with a fraction of $(7.11\pm1.03)\%$ at most.
For comparison, the RR, RN and SC galaxies are plotted in both 
figures. Most of the RR galaxies fall in the cyan areas as 
expected. 

We conclude that, although the quenched but \hi-rich galaxies 
like those in our RR sample do exist in the nearby Universe, 
they contribute only a tiny fraction to the general population 
of their colors and SFRs. In other words, the majority of 
massive quiescent galaxies are \hi-poor, as found in many 
previous studies (see \citealt{Saintoge2022}, \citealt{Cortese2020},
and references therein).

\subsection{Origin of the RR galaxies}

The formation scenarios of the RR galaxies are not immediately 
clear based on our results. The high fraction of central 
galaxies in the RR sample as well as the relatively low 
density of the local environment are both helpful for the 
galaxies to sustain their cold gas against environmental 
effects such as ram-pressure and tidal stripping, which 
are stronger for satellite galaxies and in denser regions. 
However, the origin of the large amounts of \hi\ gas in 
these galaxies is still unclear. As mentioned, the large 
\hi-to-optical size ratios of the RR galaxies imply 
very extended \hi\ discs with low surface densities of 
\hi\ in the outskirt. It is thus natural to expect 
no or weak star formation even with a substantially large 
amount of total \hi\ mass. This is consistent with the finding 
of \citet{Lemonias2014} where the resolved \hi\ observations 
revealed low \hi\ densities in a sample of 20 \hi-rich galaxies 
with suppressed SFRs. Resolved \hi\ observations 
	from The ATLAS$^{\rm 3D}$ \hi\ survey also show that 
	about $20\%$ of the nearby ETGs outside the Virgo Cluster 
	are surrounded by large \hi\ discs or rings of low column density 
	\citep{Serra2012-ATLAS3D,Serra2014}, with signs of recent 
star formation similar to that in the outskirt of spiral galaxies 
\citep{Yildiz2015,Yildiz2017}. We have visually examined 
the optical image of our RR galaxies from the DESI Legacy 
Survey (see \autoref{fig:img} in the \autoref{app:appendixA}), finding
about half of our galaxies to present faint and extended 
ring or spiral structures in the outskirt. This fraction could be 
even larger if deeper images are available. In addition, 
some of these galaxies show misalignment between the main 
stellar body and the extended structure, which was also 
observed for half of the ATLAS$^{\rm 3D}$ ETGs that have 
large \hi\ discs/rings \citep{Serra2012-ATLAS3D,Serra2014}. 
This result is consistent with the finding of the ATLAS$^{\rm 3D}$ 
survey, both supporting the external origin of the massive \hi\ 
gas content, such as gas accretion and/or mergers.
	
The extended \hi\ disks may be supported by high angular 
momentum, which is inherited from the high spin of host 
dark matter halos as suggested by \citet{Lutz2018} for their 
\hi-extreme galaxies. \cite{Mancera-Pina2021a} and 
\cite{Mancera-Pina2021b} have recently measured the specific 
angular momentum for both stars and atomic gas in a sample 
of nearby disc galaxies, finding a strong correlation between 
the gas fraction and the specific angular momentum, with 
higher angular momenta and lower star formation efficiencies 
for galaxies of higher gas fractions. In particular, it was found 
that the \hi-extreme galaxies from \citet{Lutz2018} tightly 
followed the same scaling correlations of angular momentum
as galaxies of normal \hi\ content.
Although our RR galaxies differ 
from those previous samples in certain aspects 
as discussed above, they could also have high 
angular momentum, because of which the \hi\ gas cannot 
be efficiently converted into stars. In fact, the
angular momentum of circum-galactic medium (CGM) has 
been found to either enhance or suppress star formation 
in simulated disc galaxies \citep{WangSen2022, Lu2022}. 
The ideal of ``angular momentum quenching''
has also been advocated in empirical models of galaxy 
evolution \citep[e.g.][]{Obreschkow2016,Peng2020}, although 
the model predictions are found to not fully match 
the current \hi\ surveys such as xGASS \citep{Hardwick2022}.
The RR galaxies studied in our work represent a unique 
population useful to test such models, but for this 
purpose one would need larger samples with both resolved 
\hi\ observations and angular momentum measurements. 

We should point out that, when estimating the \hi\ sizes 
for our galaxies we have assumed the \hi\ gas is distributed in a 
regular disk as in most \hi-bearing galaxies. In fact, the only galaxy 
in our sample with resolved \hi\ observations, UGC 1382
(id=0 in Table~\ref{tbl:galaxylist}) indeed shows a regularly rotating 
\hi\ disk. However, irregular and even off-galaxy \hi\ distributions 
have been observed for a handful of \hi-bearing ETGs, e.g. the 
giant \hi\ ring around NGC 3384 and M 96 in the Leo group 
\citep{Schneider1983,Schneider1985,Schneider1989,Michel-Dansac2010} 
and the large \hi\ ring around the quiescent galaxy 
AGC 203001 \citep{Bait2020}. As can be seen from 
\autoref{fig:comparisons}, both the Leo ring and AGC 203001
meet the selection criteria of our RR sample. Therefore, it is 
possible that some fraction of our RR galaxies also show similarly 
irregular \hi\ distributions. Although its physical origin is still under 
debate, such \hi-dominated rings are likely expelled gas from
\hi-rich ETGs caused by a collision with an intruder galaxy 
\citep[e.g.]{Michel-Dansac2010, Bait2020} or tidal tripping by the 
group potential \citep[e.g.][]{Bekki2005,Serra2013,Corbelli2021}, or falling back
of the tidal tail of gas-rich major mergers \citep[e.g.][]{Morganti2003,Sameer2022}.
Considering the extreme paucity of such observational cases in the 
literature as well as the fact that the majority of our RR galaxies have 
no close companions, we would expect such an irregular or off-galaxy 
\hi\ distribution to be not a common feature in the RR galaxies. 
Resolved \hi\ observations for the RR galaxies would be necessary
in order to test out this conjecture.

\section{Summary}\label{sec:summary}

In this work we have identified and investigated a rare population 
of {\em red but \hi-rich} (RR) galaxies in the local Universe. 
We make use of three \hi\ surveys including 
ALFALFA, xGASS and \hi-MaNGA, from which we select a
sample of 47 RR galaxies on the diagram of $\log_{10}($\mhi/\mstar$)$ 
versus \nuvr. Our RR galaxies are required to have \nuvr~$>5$,
and be outliers in the distribution of $\log_{10}($\mhi/\mstar$)$,
located outside the 95\% contour of a volume-limited sample of 
SDSS for which we have estimated an \hi\ mass for each galaxy 
based on Bayesian inferences of its optical properties. 
The RR galaxies are relatively massive, 
with stellar mass \mstar$~\ga 10^{10}~$\msun. For comparison
we have selected a control sample of ``red and \hi-normal''
(RN) galaxies that are closely matched with the RR sample 
in \mstar\ and \nuvr, as well as a ``stellar mass control''
(SC) sample that is matched only in \mstar. We have demonstrated 
	that the RN sample is representative to the massive red galaxy population (\autoref{fig:diagrams} and \autoref{fig:clustering}).

We have examined a variety of optical properties, finding the 
RR and RN samples to be very similar in all the properties, 
which are typical of massive red (quenched) galaxies in the 
SDSS volume-limited sample. We have also examined the environment 
of our samples, as quantified by three different statistics:
overdensity of the local environment $(1+\delta)$ as 
estimated at 2 Mpc and 4 Mpc,  projected two-point 
cross-correlation 
function $w_p(r_p)$ with respect to a reference sample of 
half a million galaxies from the SDSS spectroscopic survey, 
and background-subtracted neighbor 
counts $N_c(<R_p)$ estimated with the SDSS photometric sample
down to $r$-band magnitude of $r_{\rm lim}=21$ mag. We find the RR 
sample to be associated with lower density environments at 
a scale of 4 Mpc, and have lower clustering amplitudes and 
smaller neighbor counts at intermediate scales from a few 
$\times100$kpc to a few Mpc, when compared to the RN sample. 
These results imply that the RR galaxies are preferentially 
located at the center of relatively low-mass dark matter halos. 
This is confirmed by the SDSS galaxy group catalog which reveals 
that about 89\% of our 
RR galaxies are indeed the central galaxy of their groups, 
compared to a central fraction of $\sim$60\% for the RN sample.
The median halo mass of the RR sample is found to be 
$M_h\sim10^{12}h^{-1}\msun$, compared to $M_h\sim10^{12.5}h^{-1}\msun$ 
for the RN sample. 
Furthermore, we have estimated the \hi\ disk size for each 
of our galaxies based on the tight \hi\ size-mass relation
from the literature. We find an average \hi-to-optical size 
ratio of $R_{\rm HI}/R_{90}\sim4$ for the RR sample, which 
is four times the average ratio found for the RN sample. 

We have compared our RR sample with similar samples investigated
in previous studies. We find the previous samples are dominated by star-forming galaxies with extremely-high \hi\ 
content, or quenched \hi-poor galaxies, or transition galaxies with blue-to-green colors and 
intermediate SFRs, thus falling in between the star-forming 
main sequence and the quiescent sequence. Our RR galaxies are 
unique for their (NUV-to-optical) red colors indicative of
fully-quenched star formation status and simultaneously 
high fractions of \hi\ mass. We use the SDSS volume-limited 
sample with estimated \hi\ masses to show that, although 
the RR galaxies form a unique population, they are rather rare, 
contributing only a tiny fraction of the quenched galaxy
population. The majority of massive quenched galaxies in 
the low-redshift Universe are \hi-poor.

\section*{Acknowledgments}

We are grateful to the anonymous referee for the helpful comments.
This work is supported by the National Key R\&D Program of China
(grant NO. 2022YFA1602902), and the National Natural Science 
Foundation of China (grant Nos. 11821303, 11733002, 11973030, 
11673015, 11733004, 11761131004, 11761141012). 
This work has made use of the following softwares: 
Numpy \citep{2020Natur.585..357H},
Matplotlib \citep{2007CSE.....9...90H}, 
Astropy \citep{Astropy2013,Astropy2018,Astropy2022}, Scipy \citep{2020SciPy-NMeth}, corner \citep{corner}, and Seaborn \citep{seaborn}.

Funding for the SDSS and SDSS-II has been provided by the Alfred P. Sloan Foundation, the Participating Institutions, the National Science Foundation, the U.S. Department of Energy, the National Aeronautics and Space Administration, the Japanese Monbukagakusho, the Max Planck Society, and the Higher Education Funding Council for England. The SDSS Web Site is http://www.sdss.org/.

The SDSS is managed by the Astrophysical Research Consortium for the Participating Institutions. The Participating Institutions are the American Museum of Natural History, Astrophysical Institute Potsdam, University of Basel, University of Cambridge, Case Western Reserve University, University of Chicago, Drexel University, Fermilab, the Institute for Advanced Study, the Japan Participation Group, Johns Hopkins University, the Joint Institute for Nuclear Astrophysics, the Kavli Institute for Particle Astrophysics and Cosmology, the Korean Scientist Group, the Chinese Academy of Sciences (LAMOST), Los Alamos National Laboratory, the Max-Planck-Institute for Astronomy (MPIA), the Max-Planck-Institute for Astrophysics (MPA), New Mexico State University, Ohio State University, University of Pittsburgh, University of Portsmouth, Princeton University, the United States Naval Observatory, and the University of Washington.

\bibliography{cite_list}{}
\bibliographystyle{aasjournal}

\appendix 

\section{The optical images of the RR sample} \label{app:appendixA}

\autoref{fig:img} displays the optical images of the RR sample. The image of galaxy 39 is taken from the SDSS DR16 Skyserver \footnote{https://skyserver.sdss.org/dr16/en/home.aspx}, and the images of the other galaxies are taken from the DESI Legacy Survey \footnote{https://www.legacysurvey.org}. The images are ordered by decreasing \hi-to-stellar mass ratio from top left to bottom right.
The ID and \hi-to-stellar mass ratio of each galaxy are indicated on top of the corresponding image.

\begin{figure*}[ht!]
    \centering
    \includegraphics[width=\textwidth]{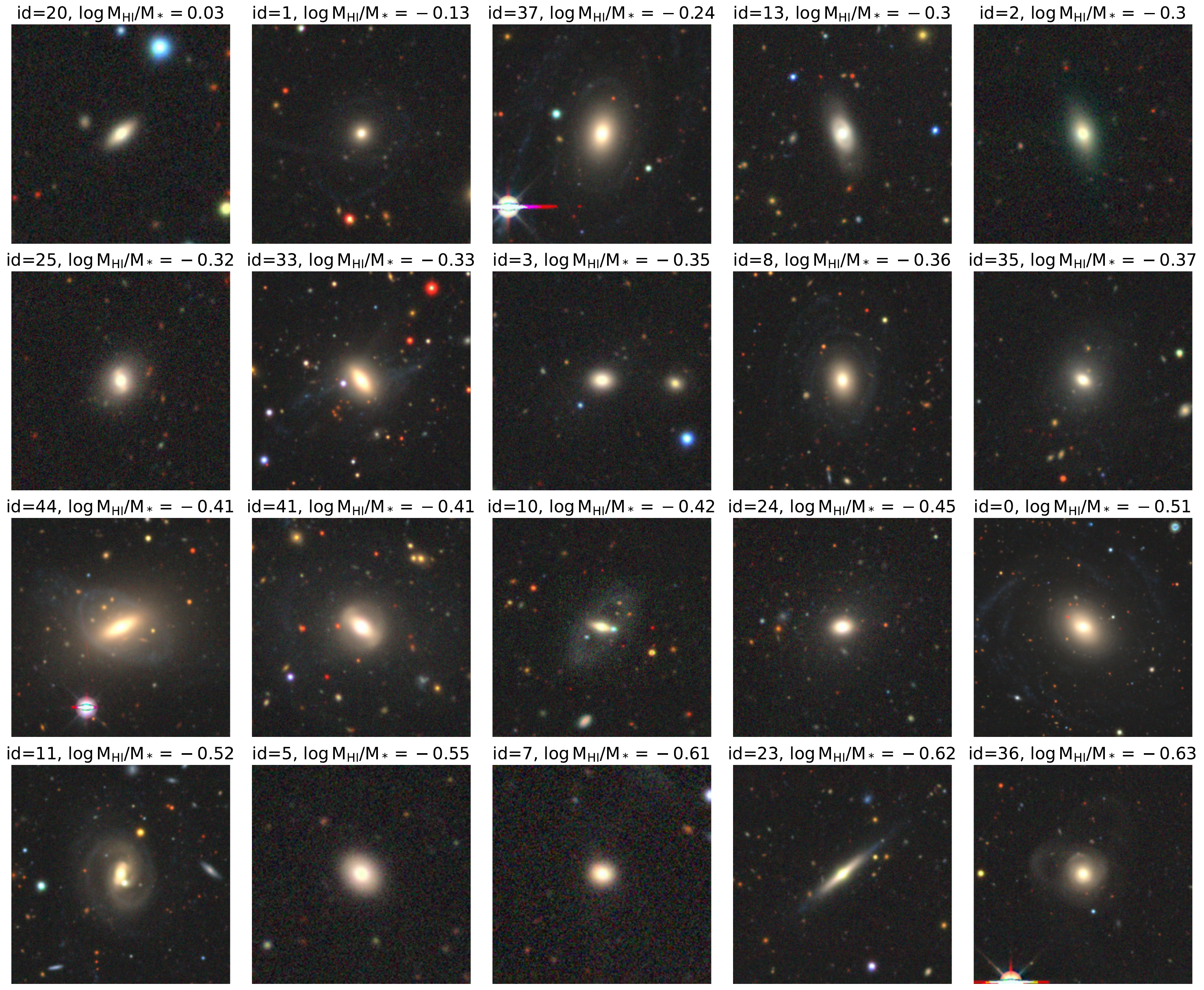}
    \caption{The optical images of the RR galaxies. The galaxies are ordered by decreasing \hi-to-stellar mass ratio ($\log_{10}\mathrm{M_{HI}/M_{\ast}}$). 
    	The galaxy ID from Table~\ref{tbl:galaxylist} and the value of $\log_{10}\mathrm{M_{HI}/M_{\ast}}$ for each galaxy are indicated on top of the corresponding image.}
    \label{fig:img}
\end{figure*}

\addtocounter{figure}{-1}
\begin{figure*}
    \centering
    \includegraphics[width=\textwidth]{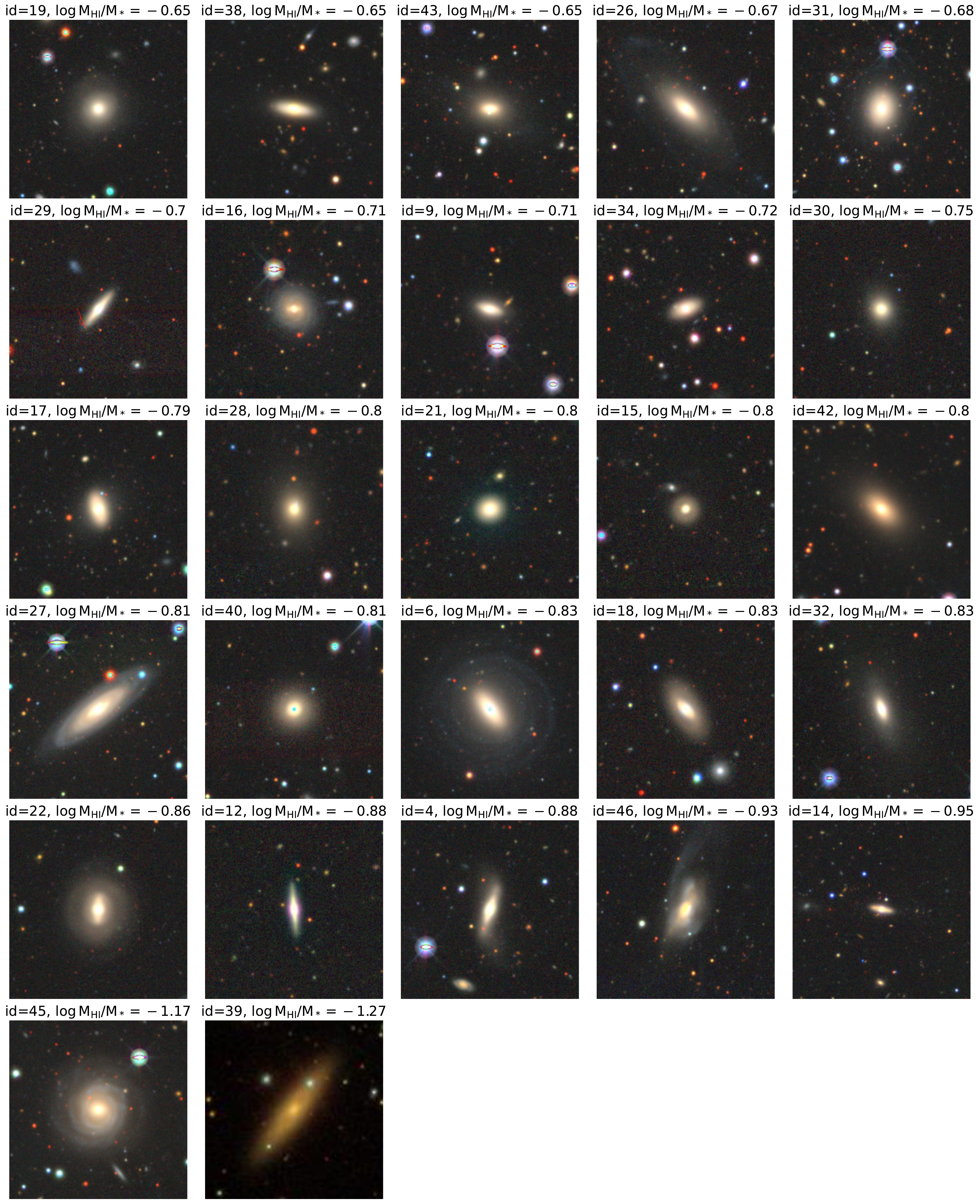}
    \caption{Continued.}
\end{figure*}

\end{document}